\documentclass[UTF8, twocolumn, times]{aastex63}
\hypersetup{linkcolor=blue,citecolor=blue,filecolor=blue,urlcolor=blue}

\usepackage{graphicx}
\usepackage{array}
\usepackage{booktabs}
\usepackage{float}
\usepackage{color}
\usepackage{CJKutf8}

\usepackage{enumitem}
\setlist[itemize]{leftmargin=*}
\turnoffedit

\begin{document}
\begin{CJK*}{UTF8}{gbsn}
	
	\title{Exploring the Link between the X-ray Power Spectra and Energy Spectra of Active Galactic Nuclei}
	
	\correspondingauthor{Haonan Yang, Chichuan Jin}
	\email{hnyang@nao.cas.cn, ccjin@nao.cas.cn}
	
	\author{Haonan Yang (杨浩楠)}
	\affiliation{National Astronomical Observatories, Chinese Academy of Sciences, 20A Datun Road, Beijing 100101, China}
	\affiliation{School of Astronomy and Space Sciences, University of Chinese Academy of Sciences, 19A Yuquan Road, Beijing 100049, China}
	
	\author{Chichuan Jin (金驰川)}
	\affiliation{National Astronomical Observatories, Chinese Academy of Sciences, 20A Datun Road, Beijing 100101, China}
	\affiliation{School of Astronomy and Space Sciences, University of Chinese Academy of Sciences, 19A Yuquan Road, Beijing 100049, China}
	
	\author{Weimin Yuan}
	\affiliation{National Astronomical Observatories, Chinese Academy of Sciences, 20A Datun Road, Beijing 100101, China}
	\affiliation{School of Astronomy and Space Sciences, University of Chinese Academy of Sciences, 19A Yuquan Road, Beijing 100049, China}

	\begin{abstract}
		
		Active Galactic Nuclei (AGN) are generally considered as the scaled-up counterparts of X-ray binaries (XRBs). It is known that the power spectral density (PSD) of the X-ray emission of XRBs shows significant evolution with spectral states. It is not clear whether AGN follow a similar evolutionary trend, however, though their X-ray emission and the PSD are both variable. In this work, we study a sample of nine AGN with multiple long observations with \textit{XMM-Newton}, which exhibit \edit1{{significant}} X-ray spectral variation. We perform Bayesian PSD analysis to measure the PSD shape and variation. We find that a large change in the X-ray energy spectrum (mainly the change of flux state) is often accompanied by a large change in the PSD shape. The emergence of a high-frequency break in the PSD also depends on the spectral state. Among the four sources with significant high-frequency PSD breaks detected, three show the break only in the high-flux state, while the remaining one shows it only in the low-flux state. Moreover, the X-ray rms variability in different spectral states of an AGN is found to vary by as much as 1.0 dex. These results suggest that the different variability properties observed are likely caused by different physical processes dominating different spectral states. \edit1{{Our results also indicate that the intrinsic PSD variation can introduce a significant fraction of the dispersion as reported for the correlations between various X-ray variability properties and the black hole mass.}} 
		
	\end{abstract}
	

\section{Introduction} \label{sec:1}
	
	Active Galactic Nuclei (AGN) and black hole X-ray binaries (XRBs) are both powered by gas accretion on to central black holes. Their observational properties are thought to be determined mainly by the fundamental parameters of black hole, such as mass and spin, together with the mass accretion rate. AGN and XRBs are both X-ray emitters. As their X-rays are believed to be predominately emitted from a region close to the black hole, the study of X-rays can provide a wealth of information about the black hole and its surroundings. 
	In the case of XRBs, the X-rays come mainly from the disc and the corona, and the relative dominance depends on which of the spectral/accretion states the system is actually in. Particularly, the low/hard state shows hard spectrum with a power law shape thought to arise from Comptonized emission, while in the high/soft state the X-ray flux is dominated by thermal radiation from the accretion disc; and the `very high state' has very high flux together with a very soft spectrum (see the review by \citealt{Remillard2006}). In contrast, the X-ray emission of AGN is thought to be mainly originated from the hot corona above the accretion disc, while the disc itself has a much lower temperature and mainly radiates in the optical and UV bands.

	A long-standing idea is that AGN are the scaled-up counterparts of XRBs, and the accretion processes in the two systems are comparable. In the past decades, many studies have been devoted to the comparative studies of AGN and XRBs. For instance, \citet{Merloni2003} suggested a common `fundamental plane' formed by the radio and X-ray luminosities of AGN and XRBs. \citet{McHardy06} proposed an anti-correlation between the high-frequency break of the X-ray power spectral density (PSD) and the black hole mass which extends from XRBs to AGN. \citet{McHardy2007b} reported two Lorentzian components in the broadband X-ray PSD of the narrow-line Seyfert 1 (NLS1) galaxy Ark 564, similar to the PSD of XRBs which also consists of multiple Lorentzian components (e.g. \citealt{Axelsson2005}). \citet{Zhou2015} found that the frequencies of quasi-periodic oscillation (QPO) has a good inverse scaling relation with the black hole mass extending from XRBs to AGN. \citet{Jin2021} reported that the X-ray QPO signal in the NLS1 RE J1034+396 has a similar rms spectrum and phase-lag spectrum as the 67 Hz QPO of the black hole XRB GRS 1915+105. Moreover, it was reported that there exists a linear rms-flux correlation in the X-ray light curves of both XRB and AGN (e.g. \citealt{Gaskell2004}; \citealt{Uttley2005}). The correlation between the hard X-ray photon index and the Eddington ratio was also found to be similar between AGN and XRBs (\citealt{Yang2015}). Therefore, although the two types of accretion systems have masses differing by a factor of $10^{5-9}$, many similar properties have been found between them.
	
	Our current understanding about the state transition of XRBs is much deeper than AGN. An important observation is the co-evolution of the energy spectrum and PSD of XRBs during the transition, which is crucial for understanding the geometry of disc-corona and its evolution with the accretion rate. It is known that the X-ray PSD of XRBs is not stationary, but evolves with the accretion state. As the spectral state evolves from the hard state to the soft, the components in the PSD all move towards higher frequencies, and eventually disappears at a roughly constant frequency (e.g. \citealt{Axelsson2005}; \citealt{Gierlinski2008}). This is qualitatively well explained by the evolution of truncated disc model (see the review by \citealt{Done2007}).
	
	In comparison, the relationship between the X-ray spectra and PSD evolution of AGN is still unclear. The variability timescale of AGN is at least $10^{5}$ longer than XRBs, so it is more difficult to monitor the state transition of AGN. Recent attempts mainly focused on the rapid spectral variation of rare changing-look AGN (e.g. \citealt{Lamassa2015}). For instance, \citet{Noda2018} studied the evolution of broadband spectral energy distribution (SED) of the AGN Mrk 1018, and found that its SED evolution was similar to XRBs, although the effect of radiation pressure may be more important in AGN. \citet{Ruan2019b} collected a sample of changing-look AGN, and discovered that the correlation formed by their UV-to-X-ray spectral indices and Eddington ratios is remarkably similar to the state transition of XRBs.
	
	However, there have been few studies on the link between the X-ray spectrum and PSD of normal AGN. It has been known that the PSDs of some AGN are variable. For instance, \citet{Jin2021} found that the X-ray PSD of RE J1034+396 does not show a QPO signal when the soft X-ray excess becomes stronger. \citet{Gonzalez-Martin2018} reported a correlation between the X-ray absorption column density ($N_{\rm H}$) and the high-frequency break in the PSD of AGN. Since they only used the 2-6 keV spectrum to determine $N_{\rm H}$, their result may also be ascribed to a correlation between the observed X-ray spectral slope and the break frequency, i.e. the harder the spectrum, the higher the break frequency is.
	
	It is also worth noting that the light curve observed in a single observation of AGN is only a random realization of the intrinsic variability (\citealt{Timmer1995}). Hence even if the intrinsic PSD is stationary, the observed periodogram, which is a realization of the intrinsic power spectrum, would still vary from one observation to another (e.g. \citealt{Vaughan2005, Vaughan2010}). The X-ray rms amplitude can vary significantly in different segments of a single light curve from a stationary process (\citealt{Vaughan2003b}). Therefore, when comparing two PSDs or quantifying PSD evolution, one must consider random fluctuation properly. \citet{Vaughan2010} proposed a Bayesian test for the existence of QPO signal in the PSD of AGN (also see \citealt{Jin2020}). In this work, we generalize this method to measure and compare PSD shapes.
	
	Moreover, the variation of spectral properties such as the flux and slope generally reflect intrinsic variation of physical processes and parameters. Therefore, we use a sample of radio-quiet AGN to explore the link between the variability of their X-ray spectra and PSDs. The choice of radio-quietness is to avoid complexities due to the jet emission. Since the AGN in this sample all exhibit \edit1{{significant}} spectral variation, their PSD variation is more likely to be intrinsic rather than due to random fluctuation. We will show that there is indeed evidence to support the existence of such a link.
	
	This paper is organized as follows. Firstly we describe the AGN sample, observations and data reduction in Section \ref{sec:2}. Then we describe the methodology of our spectral fitting and PSD analyses in Section \ref{sec:3}. This is followed by the investigation of potential link between the state of energy spectra and PSDs in Section \ref{sec:4}. In Section \ref{sec:5} we discuss the variability of PSD and the possible implications and explanations, as well as the impact of PSD variation on various correlations used to estimate the black hole mass. Section \ref{sec:6} summarizes the results of this work.

	\begin{table*} 
		\renewcommand\arraystretch{1.2}
		\begin{center}
			\caption{List of sources and \textit{XMM-Newton} observations used in this paper. \label{tab:obs}}
			\setlength\tabcolsep{4.0pt}
			\begin{tabular}{@{}llcccccccc@{}}
				\hline
	Source Name            & Type  & Obs ID     & Date    & Duration & Pile-up & bin & $\Gamma_{\rm 2-10 keV}$     & $F_{\rm 0.3-1~keV}$       & $F_{\rm 2-10~keV}$       \\
				&       &            &         & (ks)   & ($''$)  & (s) &  & (erg cm$^{-2}$ s$^{-1}$)  & (erg cm$^{-2}$ s$^{-1}$)\\ \hline
				Mrk 335         & NLS1 & 0306870101 & 2006-01 & 110.4 &    --     & 50  & $2.05\pm 0.01$   & 2.27$\times 10^{-11}$    & 1.85$\times 10^{-11}$   \\
				&       & 0600540601 & 2009-06(1) & 100.0  & 7.5     & 50  & $1.48\pm 0.02$   & 2.76$\times 10^{-12}$   & 5.50$\times 10^{-12}$   \\
				&       & 0600540501 & 2009-06(2) & 80.7   & 9.0      & 50  & $1.65\pm 0.02$   & 3.91$\times 10^{-12}$   & 5.20$\times 10^{-12}$   \\ \hline
				Mrk 766         & NLS1 & 0109141301 & 2001-05 & 80.1   & 2.5     & 50  & $2.05\pm 0.01$   & 2.55$\times 10^{-11}$    & 2.61$\times 10^{-11}$   \\
				&       & 0304030101 & 2005-05(1) & 70.1  & 1.5     & 50  & $1.31\pm 0.01$   & 4.45$\times 10^{-12}$  & 7.63$\times 10^{-12}$   \\
				&       & 0304030501 & 2005-05(2) & 70.1  & 2.5     & 50  & $1.95\pm 0.01$   & 1.33$\times 10^{-11}$   & 1.75$\times 10^{-11}$   \\ \hline
				1H 0707-495     & NLS1 & 0148010301 & 2002-10 & 76.4   & 10.0      & 200 & $2.27\pm 0.05$   & 4.71$\times 10^{-12}$   & 1.25$\times 10^{-12}$   \\
				&       & 0506200301 & 2007-05 & 38.6   & 10.0      & 200 & $1.94\pm 0.10$   & 2.53$\times 10^{-12}$   & 6.51$\times 10^{-13}$   \\
				&       & 0653510301 & 2010-09 & 99.8   & 13.0      & 200 & $2.08\pm 0.06$   & 5.01$\times 10^{-12}$   & 1.02$\times 10^{-12}$   \\ \hline
				IRAS 13224-3809 & NLS1  & 0110890101 & 2002-01 & 35.0     & 10.5    & 200 & $1.66\pm 0.13$   & 2.31$\times 10^{-12}$   & 4.10$\times 10^{-13}$ \\
				&       & 0673580401 & 2011-07 & 109.8  & 11.0     & 200 & $2.20\pm 0.07$   & 2.62$\times 10^{-12}$   & 4.72$\times 10^{-13}$ \\
				&       & 0780561301 & 2016-07 & 119.8  & 11.0      & 200 & $2.23\pm 0.06$   & 3.25$\times 10^{-12}$   & 4.35$\times 10^{-13}$ \\ \hline
				NGC 1365        & S1.8  & 0505140401 & 2007-07 & 120.1  & 1.0       & 50  & $-0.29\pm 0.03$  & 4.66$\times 10^{-13}$   & 1.89$\times 10^{-12}$   \\
				&       & 0692840201 & 2012-07 & 100.0  & 1.0       & 50  & $-0.52\pm 0.01$  & 4.31$\times 10^{-13}$   & 7.81$\times 10^{-12}$  \\
				&       & 0692840501 & 2013-02 & 113.3 & 0.5     & 50  & $0.10\pm 0.01$   & 4.69$\times 10^{-13}$   & 1.34$\times 10^{-11}$   \\ \hline
				NGC 3516        & S1.5  & 0107460701 & 2001-11 & 80.0  & 3.0       & 50  & $0.87\pm 0.01$   & 1.60$\times 10^{-12}$   & 1.66$\times 10^{-11}$   \\
				&       & 0401210601 & 2006-10 & 50.1   & 1.0       & 50  & $1.36\pm 0.01$   & 7.15$\times 10^{-12}$   & 3.52$\times 10^{-11}$   \\ \hline
				NGC 4051        & NLS1 & 0109141401 & 2001-05 & 101.4  & 2.5     & 50  & $1.85\pm 0.01$   & 2.86$\times 10^{-11}$   & 2.49$\times 10^{-11}$  \\
				&       & 0830430801 & 2018-11 & 42.8   & 1.0       & 50  & $1.65\pm 0.01$   & 8.67$\times 10^{-12}$   & 1.20$\times 10^{-11}$   \\ \hline
				NGC 4395        & S1.8  & 0142830101 & 2003-11 & 89.9   & 6.0       & 50 & $0.98\pm 0.01$   & 4.44$\times 10^{-13}$   & 5.92$\times 10^{-12}$   \\
				&       & 0744010101 & 2014-12 & 51.8   &    --     & 50 & $0.003\pm 0.021$ & 4.23$\times 10^{-14}$ & 4.66$\times 10^{-12}$   \\
				&       & 0824610401 & 2019-01 & 74.8   &     --    & 50 & $0.78\pm 0.01$   & 5.27$\times 10^{-14}$ & 5.09$\times 10^{-12}$   \\ \hline
				PG 1211+143     & NLS1 & 0112610101 & 2001-06 & 53.1   & 6.0       & 50  & $1.72\pm 0.03$   & 3.86$\times 10^{-12}$   & 3.16$\times 10^{-12}$   \\
				&       & 0502050101 & 2007-12 & 45.1   & 9.0       & 50  & $2.03\pm 0.03$   & 7.69$\times 10^{-12}$   & 4.01$\times 10^{-12}$   \\ \hline
			\end{tabular}
		\end{center}
		\vspace{-0.3cm}
		\tablecomments{Type: AGN classifications from \citet{Veron-Cetty2010} and \citet{Alston2019}; Duration: total length of light curve used for variability analysis from each observation; Pile-up: radii of removed central area for suppressing photon pile-up; bin: time bins of different light curves; $\Gamma_{\rm 2-10~keV}$: 2-10 keV photon indices with 90\% confidence intervals derived from {\sc xspec} fitting; $F_{\rm 0.3-1~keV}$ and $F_{\rm 2-10~keV}$: the fluxes in 0.3-1 and 2-10 keV obtained by integrating the energy spectrum.}
		\vspace{0.6cm}
	\end{table*}

\section{Sample Selection and Data Reduction} \label{sec:2}
	
	\subsection{The AGN Sample}
	The AGN sample of this work is based on the larger sample in \citet{Kara2016b} which, in turn, comes from a sample of 104 variable AGN presented originally by \citet{Gonzalez-Martin2012}. The merit of \citet{Kara2016b}'s sample is that those AGN are strongly variable in X-rays, so that possible evolution of X-ray timing properties can be investigated. Since our objective is to explore the link between the X-ray energy spectrum and PSD, we inspected the spectra of each source in \citet{Kara2016b} to identify those with multiple \textit{XMM-Newton} observations and exhibiting \edit1{{significant}} spectral variation, i.e. the 2-10 keV photon indices differ by more than 0.2 between two observations of the same source. Then we select sources having at least 2 \textit{XMM-Newton} observations of more than 40 ks each, which ensures that their PSDs can be compared down to $\sim10^{-5}$ Hz. Moreover, we only use observations which are not severely affected by background flares. The final sample consists of nine highly variable AGN with 24 \textit{XMM-Newton} observations. Details of these observations are listed in Table \ref{tab:obs}, where the AGN classification comes from various literatures (e.g. \citealt{Veron-Cetty2010}, \citealt{Alston2019}). The sample consists of one Seyfert 1.5 galaxy, two Seyfert 1.8 galaxies and six narrow-line Seyfert 1 galaxies (NLS1s). \edit1{{According to the AGN unified model, the difference between Seyfert 1.5, 1.8 and 1 is mainly the difference in viewing angle \citep{Antonucci1993, Urry1995}, and their X-ray variations may all originate from intrinsic changes in the nuclear region and/or the absorption in the line of sight \citep{Hernandez-Garcia2017}, so we can study them in one sample and compare their properties.}}
	
	\subsection{Data Reduction}
	We use the data from the European Photon Imaging Camera (EPIC) pn camera of \textit{XMM-Newton} as its data quality is the best among the three EPIC cameras. The data were processed using the \textit{XMM-Newton} Science Analysis System (SAS v. 17.0.0) with the latest calibration files. The calibrated EPIC event files were produced from the Observation Data Files (ODFs) using the {\tt \string epproc} command. For each source, we extracted a single-event 10-12 keV light curve from a source-free region to identify intervals of background flares. Generally, this was done by removing the high background periods at the beginning and end of the observation data, while the {\tt \string RATE\textless=0.4} command was also used when background flares appeared in the middle of the observing window. We created a good-time-interval (GTI) file, and filtered the event file to include data only in the low-background periods. Only single and double pixel events were used (i.e. {\tt \string PATTERN\textless=4}) in the following analysis.
	
	For most observations, we used 50 s as the binning time. For observations when the source appears relatively faint (i.e. in observations of 1H 0707-495 and IRAS 13224-3809), we adjusted the binning time to 200 s to reduce the number of zero-count bins in the light curves. The source and background light curves and spectra were extracted from the filtered event file using the {\tt \string evselect} command. In most cases, the radii of the source and background regions were 40 arcsec and 50 arcsec, respectively. Only if the observation mode was {\tt Small Window} (e.g. the observation of Mrk 335 in 2006), the two radii were both set as 20 arcsec.  We also checked every observation for the pile-up effect \edit1{{using {\tt\string epatplot} task}}, and the core area of the point-spread-function (PSF) was masked when it is necessary to reduce pile-up. \edit1{{We make sure that the observed pattern distributions as calculated by the {\tt\string epatplot} task are consistent with the models, and the energy fraction lose is less than 3.5\% in all the observations.}} The radii of the masked inner regions are listed in Table \ref{tab:obs}. The redistribution matrix files (RMFs) and ancillary files (ARFs) are generated using the {\tt\string rmfgen} and {\tt\string arfgen} commands, respectively.

\section{Methodologies of X-ray Spectral and PSD Analysis} \label{sec:3}
	
	\subsection{Spectral Analysis}
	
	To quantify the spectral shape, the 2-10 keV spectra were fitted with an absorbed power law model (the {\sc tbabs$\times$powerlaw} in XSPEC v12.10.1). The abundances were adopted from \citet{Wilms2000}, and the photoelectric absorption cross-sections were adopted from \citet{Verner1996}. The Galactic hydrogen column density $N_{\rm H}$ of each source was frozen at the value provided by the {\tt NHtot} tool\footnote{\url{https://www.swift.ac.uk/analysis/nhtot/index.php}} (Willingale et al. 2013). The free parameters are the photon index and the normalization of the power law. The best-fit photon indices are listed in Table \ref{tab:obs}, with errors representing the 1$\sigma$ confidence intervals. Since our study focuses on the PSD, these parameters are sufficient to make a simple distinction between different spectral states.
	
	\edit3{{The folded spectra are plotted in Figures \ref{fig:spec&PSD_s} and \ref{fig:spec&PSD_h}, as well as in Figures \ref{fig:spec&PSD_unused_s} and \ref{fig:spec&PSD_unused_h}. These spectra include various instrumental features (e.g. effective area and detector's response), but they are model-independent. Meanwhile, for the ease of comparison with previous works such as \citet{Kara2016b}, we follow the prescription in \citet{Vaughan2011} to plot the `fluxed' 0.3-10 keV spectra in Figure \ref{fig:unfoldspec}. These spectra are unfolded through a power law model with photon index equal to 0 and normalization equal to 1. This method largely removes the effects of the energy-dependent effective area, so the resultant `fluxed' spectra can better reflect the difference of the intrinsic spectral states. Significant spectral variation is found for every source.}}
	
	\subsection{PSD Analysis}
	The light curves of the soft (0.3-1 keV) and hard (2-10 keV) X-ray bands were extracted, separately. For the short-period data gaps resulting from the removal of background flares, we followed the method of \citet{Ashton2021} to use linear interpolation to fill them. We performed Fast Fourier Transform (FFT) to the light curves to generate periodograms. We also compared the periodograms thus obtained with those produced by filling data gaps with the mean count rates, and found that they are not significantly different.

	Then the periodograms were fitted with two models. Model-A is a single power law model for the red noise continuum, plus a constant to account for Poisson noise:
	
	\begin{equation}
		P(f)=Af^{-\alpha}+C,
	\end{equation}	
	where the free parameters include the normalization $A$, the power law slope index $\alpha$, and the constant $C$.
	
	Model-B is a broken power law plus a constant:
	
	\begin{equation}
		P(f)=\frac{Af^{-\alpha_{\rm L}}}{1+(f/f_{\rm B})^{\alpha_{\rm H}-\alpha_{\rm L}}}+C,
	\end{equation}
	where the slope below the break frequency $f_{\rm B}$ is fixed as $\alpha_{\rm L}=1$,  leaving only four free parameters: the normalization $A$, the break frequency $f_{\rm B}$, the slope $\alpha_{\rm H}$ after break, and the constant $C$.
	
	We fitted the periodograms using the method described in \citet{Vaughan2010} with the two models using the maximum likelihood estimation. \edit1{{Note that this method avoid the necessity of binning the periodogram to achieve certain level of signal-to-noise.}}  For a given model, the fitting process is equivalent to minimizing the following deviance function:
	
	\begin{equation}
		D=2\sum_{j=1}^{N/2}\{\frac{I_j}{S_j}+\log S_j\}
	\end{equation}
	which is twice the minus logarithm of the likelihood. $I_j$ and $S_j$ are the periodogram data and the model PSD at frequency $f_j$, respectively. The summation was performed on the entire frequency range below the Nyquist frequency. To obtain the confidence limits of the PSD parameters, we follow the Bayesian analysis method detailed in \citet{Vaughan2010} and use the Markov Chain Monte Carlo (MCMC) method to obtain posterior distributions of the model parameters. We use 100 walkers and each goes more than 1000 steps to ensure that the samples reach burn-in period. The median value of the posterior distribution is used as the fitting result of each parameter.

	\begin{table}[t!] 
		\centering
			\caption{LRT1 \textit{p}-values calculated from 5000 datasets simulated under Model-A. {\it soft} and {\it hard} indicate the 0.3-1 keV and 2-10 keV bands. \label{tab:LRT1}}
            \begin{tabular}{lccc}
            \hline
            Source          & Obs     & {\it p} (soft) & {\it p} (hard) \\ \hline
            Mrk 335         & 2006    & 0.0000   & 0.0000   \\
                            & 2009(1) & 0.0352   & 0.2542   \\
                            & 2009(2) & 0.0456   & 0.8408   \\
            Mrk 766         & 2001    & 0.0000   & 0.0006   \\
                            & 2005(1) & 0.0998   & 0.8642   \\
                            & 2005(2) & 0.0000   & 0.0000   \\
            1H 0707-495     & 2002    & 0.0022   & 0.0144   \\
                            & 2007    & 0.0494   & 0.0398   \\
                            & 2010    & 0.0006   & 0.0024   \\
            IRAS 13224-3809 & 2002    & 0.1090   & 0.0672   \\
                            & 2011    & 0.1716   & 0.8642   \\
                            & 2016    & 0.6278   & 0.0052   \\
            NGC 1365        & 2007    & 0.9446   & 0.7968   \\
                            & 2012    & 0.9354   & 0.0286   \\
                            & 2013    & 0.5166   & 0.0036   \\
            NGC 3516        & 2001    & 0.2948   & 0.3798   \\
                            & 2006    & 0.5906   & 0.8336   \\
            NGC 4051        & 2001    & 0.0000   & 0.0900   \\
                            & 2018    & 0.0000   & 0.0052   \\
            NGC 4395        & 2003    & 0.0178   & 0.1948   \\
                            & 2014    & 0.1918   & 0.0764   \\
                            & 2019    & 0.0194   & 0.0026   \\
            PG 1211+143     & 2001    & 0.0056   & 0.0198   \\
                            & 2007    & 0.0472   & 0.0594   \\ \hline
            \end{tabular}
		
		\vspace{-0.3cm} 
		\vspace{0.2cm} 
	\end{table}

	\subsection{Comparison of PSD models}
	
	For an observed PSD, the two models were compared using the likelihood ratio test (LRT) statistic. As described in \citet{Vaughan2010}, the calculation of the LRT statistic is equivalent to the difference between the deviance of the two models:
	
	\begin{equation}
		T_{\rm LRT}=D_{\rm min}(A) -D_{\rm min}(B).
	\end{equation}
	
	By using datasets simulated under one PSD model, the LRT statistical \textit{p}-value can be calculated, which gives the proportion of all simulated data where the LRT statistic is greater than the observed value \citep{Vaughan2010}.
	
	For the comparison of PSD models, we simulate 5000 periodograms following \citet{Timmer1995} from the posterior distribution of Model-A and calculate the LRT statistics (hereafter LRT1). Then \textit{p}-values relative to real data can be calculated. For LRT1, a smaller \textit{p}-value indicates a more significant deviation of the observed PSD from a single power law model. We take \textit{p} \textless\ 0.01 as the criterion that Model-B is selected over Model-A. \edit1{{Although this is just a crude check as in \citet{Gonzalez-Martin2012}, the selection is generally consistent with visual check result, in which the observations with \textit{p} \textless\ 0.01 do show breaks in their PSDs. Besides, considering the fact that there are only 24 observations in our sample, which is a rather limited number, this criterion is reasonable as the uncertainty implied by the \textit{p}-value would not change the result of more than one observation.}}
	
	According to the results of LRT1 statistics, four objects in our sample show \textit{p}-value lower than 0.01 in more than one dataset (different energy bands of one observation are treated as different datasets), which are Mrk 335, Mrk 766, 1H 0707-495, and NGC 4051. Besides, IRAS 13224-3805, NGC 1365, NGC 4395 and PG 1211+143 give low \textit{p}-value in only one observation (i.e. 2-10 keV in the 2016 observation for IRAS 13224-3805, 2-10 keV in the 2013 observation for NGC 1365, 2-10 keV in the 2019 observation for NGC 4395, and 0.3-1 keV in the 2001 observation for PG 1211+143). After checking the PSDs of these observations, we consider that they do not show significant break and the low LRT1 \textit{p}-values are more likely due to the power fluctuations at low frequencies.
	
	For those observations which have an LRT1 \textit{p}-value lower than 0.01 (i.e. data that can be described better by Model-B than by Model-A), MCMC gives relatively tight posterior distributions for the break frequency. These break frequencies are shown in Table \ref{tab:parameters}. \edit1{{The other PSD fit parameters of Model A and Model B are shown in Table \ref{tab:parameters_modelA} and \ref{tab:parameters_modelB}, respectively.}}

	\begin{table}[t!] 
		\begin{center}
			\caption{Break frequencies from the MCMC method. Error-bars represent 1$\sigma$ confidence intervals. \label{tab:parameters}}
			\setlength\tabcolsep{5.8pt}
			\begin{tabular}{cccc}
				\hline
				Souece      & Obs     & $\log f_{\rm B}$        & $\log f_{\rm B}$        \\
				            &         & (0.3-1 keV)             & (2-10 keV)              \\ \hline
				Mrk 335     & 2006    & $-3.88_{-0.17}^{+0.13}$ & $-3.63_{-0.13}^{+0.10}$ \\
				            & 2009(1) & --                      & --                      \\
				            & 2009(2) & --                      & --                      \\
				Mrk 766     & 2001    & $-3.38_{-0.13}^{+0.10}$ & $-3.14_{-0.24}^{+0.11}$ \\
				            & 2005(1) & --                      & --                      \\
				            & 2005(2) & $-3.31_{-0.13}^{+0.09}$ & $-3.16_{-0.06}^{+0.04}$ \\
				1H 0707-495 & 2002    & $-3.54_{-0.30}^{+0.16}$ & --                      \\
				            & 2007    & --                      & --                      \\
				            & 2010    & $-3.64_{-0.14}^{+0.10}$ & $-3.47_{-0.07}^{+0.07}$ \\
				NGC 4051    & 2001    & $-3.61_{-0.24}^{+0.18}$ & --                      \\
				            & 2018    & $-2.94_{-0.12}^{+0.09}$ & $-2.80_{-0.19}^{+0.12}$ \\ \hline
			\end{tabular}
		\end{center}
		\vspace{-0.3cm} 
		\vspace{0.4cm} 
	\end{table}

	\subsection{Comparison of Periodograms}
	\label{sec-psd-compare}
	Since a periodogram is only a single realization of the intrinsic PSD, the difference between two periodograms can be attributed to either the random fluctuation or the intrinsic PSD variation, and so it is not trivial to tell if two periodograms are intrinsically different. However, it is known that the probability distribution for the random fluctuation of PSD follows the $\chi^2$ distribution of two degrees of freedom (\citealt{Timmer1995}). Besides, the posterior probability distribution can be derived for every parameter of the PSD model (\citealt{Vaughan2010}). Hence, one can run MCMC simulation to assess the level of model uncertainty and random fluctuation, thereby identifying intrinsic PSD variation. Another issue of comparing two periodograms is the different level of Poisson noise, because if a frequency band is dominated by the Poisson noise power, then it is difficult to retrieve the PSD shape in this band. Therefore, it is also not trivial to compare two periodograms if their Poisson noise power is different. 
	
	To overcome these issues, we apply the following method to compare two periodograms. For each source, we first obtain the distributions of Model-B parameters for the observation (i.e. the reference observation) showing significant PSD break, and then adopt them as the reference \edit1{{PSD models}} for the other observations (i.e. the comparing observations) of the same source, except that the Poisson noise is kept at the level of the comparing observation. This step is to avoid the influence of different Poisson noise power between the reference and comparing observations. Similarly, \edit1{{for each comparing observation,}} we run 5000 MCMC simulations using \edit1{{Poisson noise distribution from this comparing observation and other parameters from reference observations}} in Model-B to produce a set of simulated PSDs \edit1{{based on the frequency baseline of the comparing observation}}. Then these mock PSDs are fitted again with Model-A and Model-B to derive a set of $T_{\rm LRT}$ statistics (hereafter LRT2). Based on the distribution of LRT2 statistics, the $T_{\rm LRT}^{\rm obs}$ from the real \edit1{{comparing}} observation is used to determine the \textit{p}-value. In this case, a larger \textit{p}-value can indicate that the shape of the comparing PSD is to a higher degree different from the reference PSD (i.e. the shape of broken power law).

	\begin{table}[t!] 
		\begin{center}
			\caption{LRT2 \textit{p}-value calculated from 5000 datasets simulated under Model-B. The model parameters adopt the distributions obtained by the broken PSD models of the reference observations (Ref Obs), while the Poisson terms are obtained from the comparing observations (Comp Obs). {\it soft} and {\it hard} indicate the 0.3-1 keV and 2-10 keV bands. \label{tab:LRT2}}
            \begin{tabular}{lcccc}
            \hline
            Source      & Comp Obs     & Ref Obs     & {\it p} (soft) & {\it p} (hard) \\ \hline
            Mrk 335     & 2009(1) & 2006    & 0.7876   & 0.7710   \\
                        & 2009(2) & 2006    & 0.7216   & 0.9318   \\
            Mrk 766     & 2005(1) & 2001    & 0.9218   & 0.9398   \\
                        & 2005(1) & 2005(2) & 0.9524   & 0.9860   \\
            1H 0707-495 & 2002    & 2010    & --       & 0.5224   \\
                        & 2007    & 2010    & 0.7884   & 0.5998   \\
                        & 2007    & 2002    & 0.5134   & --       \\
            NGC 4051    & 2001    & 2018    & --       & 0.9264   \\ \hline
            \end{tabular}
		\end{center}
		\vspace{-0.3cm} 
		\vspace{0.2cm} 
	\end{table}

\section{Relations between the X-ray Power Spectra and Energy Spectra} \label{sec:4}
\begin{figure*}[hp!] 
		\begin{center} 
			\gridline{
				\fig{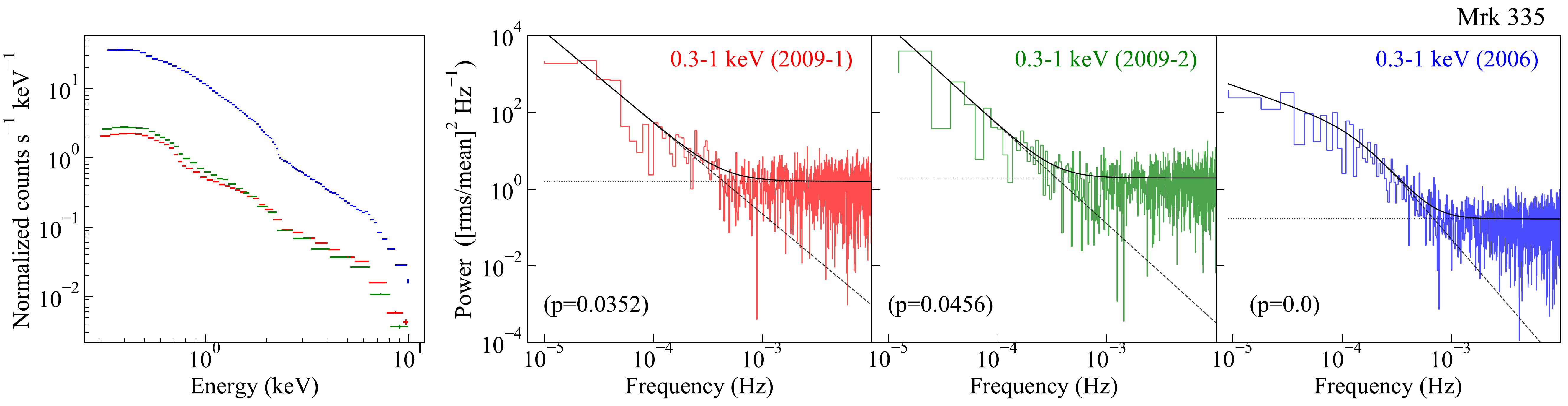}{1.0\textwidth}{}
			}
			\vspace{-0.3cm}
			\gridline{
				\fig{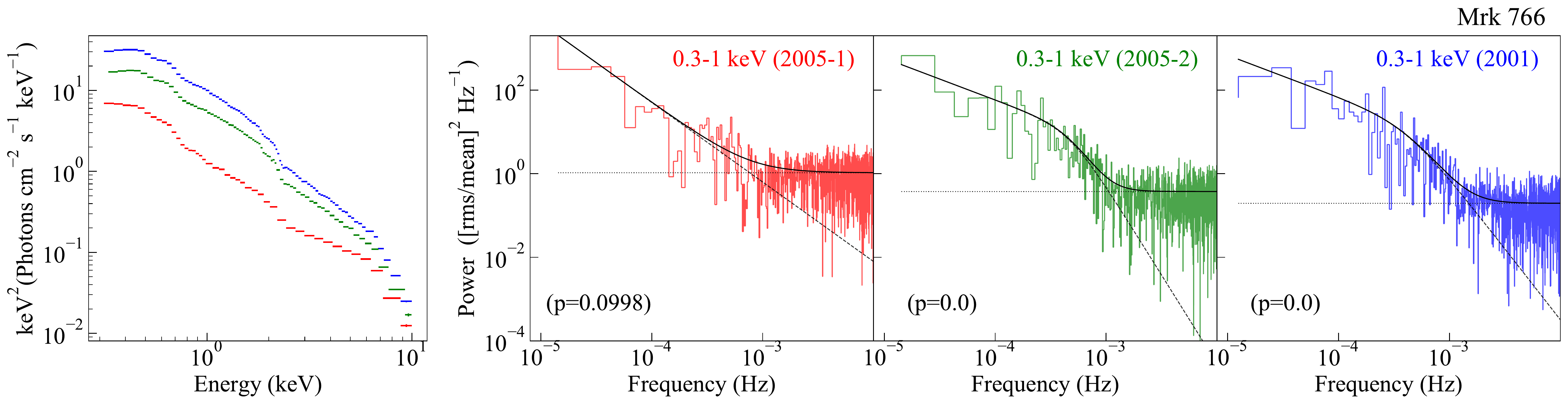}{1.0\textwidth}{}
			}
			\vspace{-0.3cm}
			\gridline{
				\fig{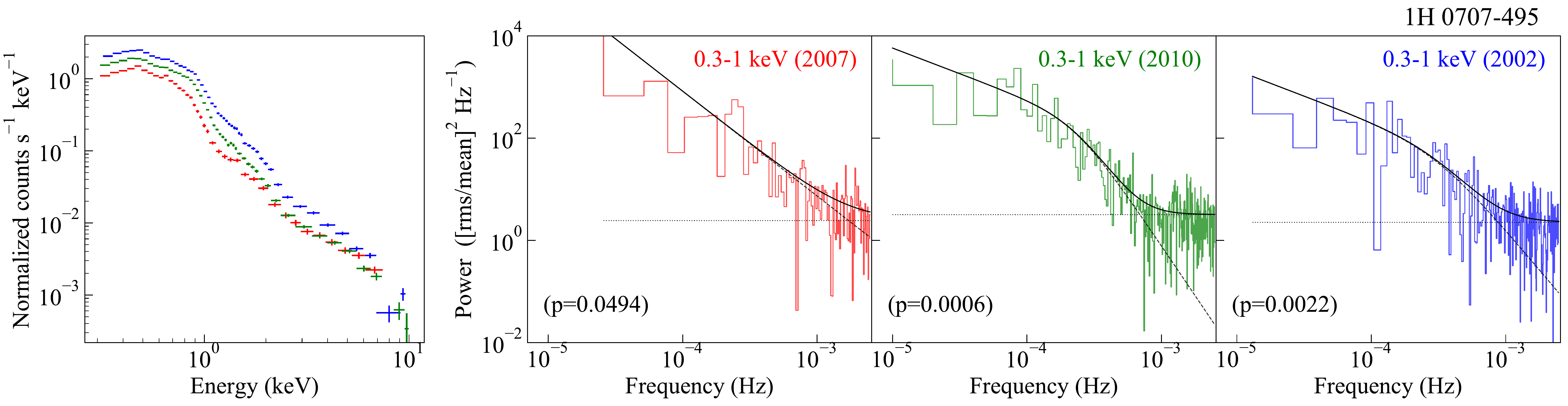}{1.0\textwidth}{}
			}
			\vspace{-0.3cm}
			\gridline{
				\fig{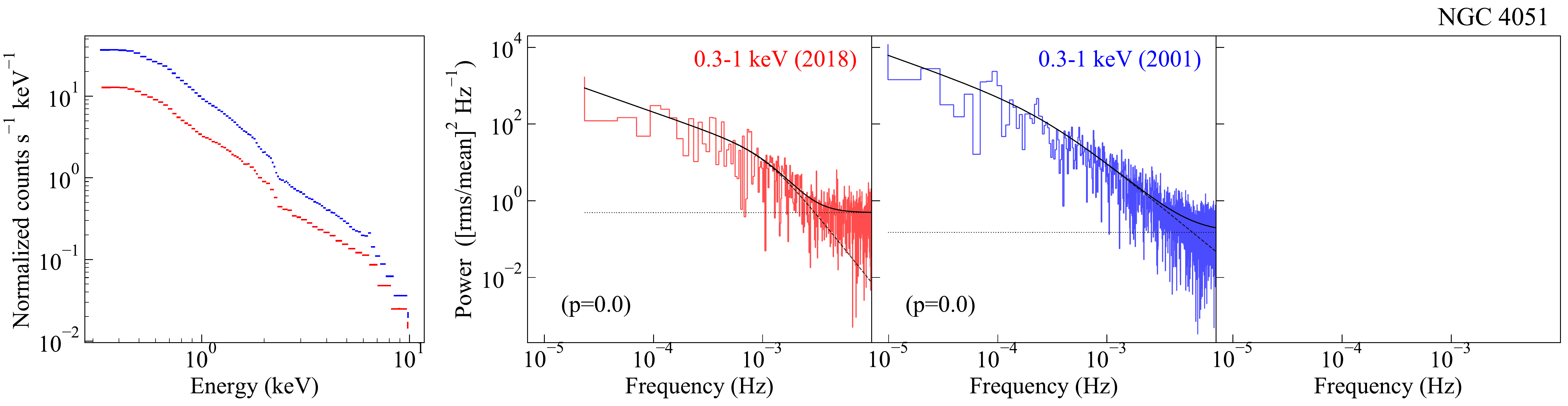}{1.0\textwidth}{}
			}
			\vspace{-0.3cm}
			\caption{The folded energy spectra and 0.3-1 keV PSDs of the four AGN which have two or more detections of high-frequency break. Black lines are the best-fit Model-B (i.e. break power law) results for observations which report significant breaks, and are the best-fit Model-A (i.e. single power law) results for observations which do not report significant breaks. The dashed lines and the dotted lines represent red noise components and Poisson terms, respectively. The LRT1 \textit{p}-values of Model-A simulation are shown downside. Spectra and PSDs are shown in different colors attending to the observation date. The PSDs are arranged in order of fluxes.  \label{fig:spec&PSD_s}}
		\end{center}
	\end{figure*}

\begin{figure*}[hp!] 
		\begin{center} 
			\gridline{
				\fig{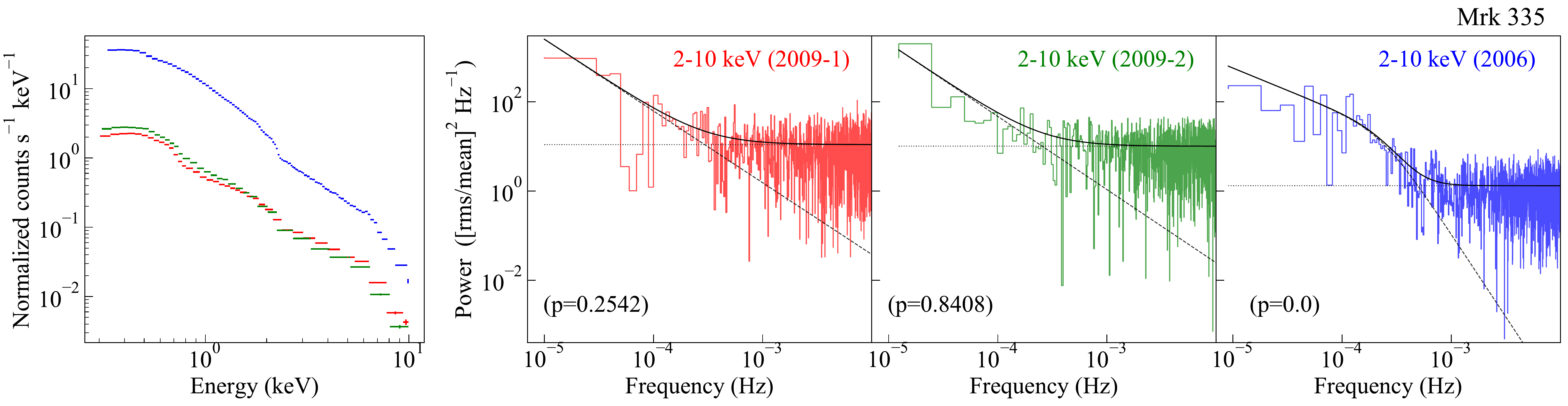}{1.0\textwidth}{}
			}
			\vspace{-0.3cm}
			\gridline{
				\fig{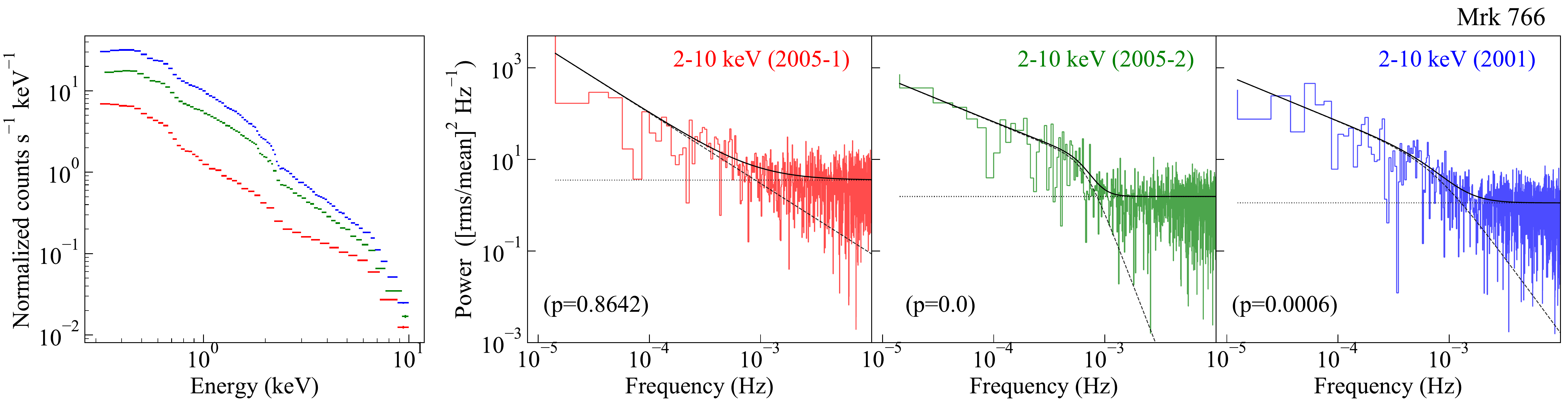}{1.0\textwidth}{}
			}
			\vspace{-0.3cm}
			\gridline{
				\fig{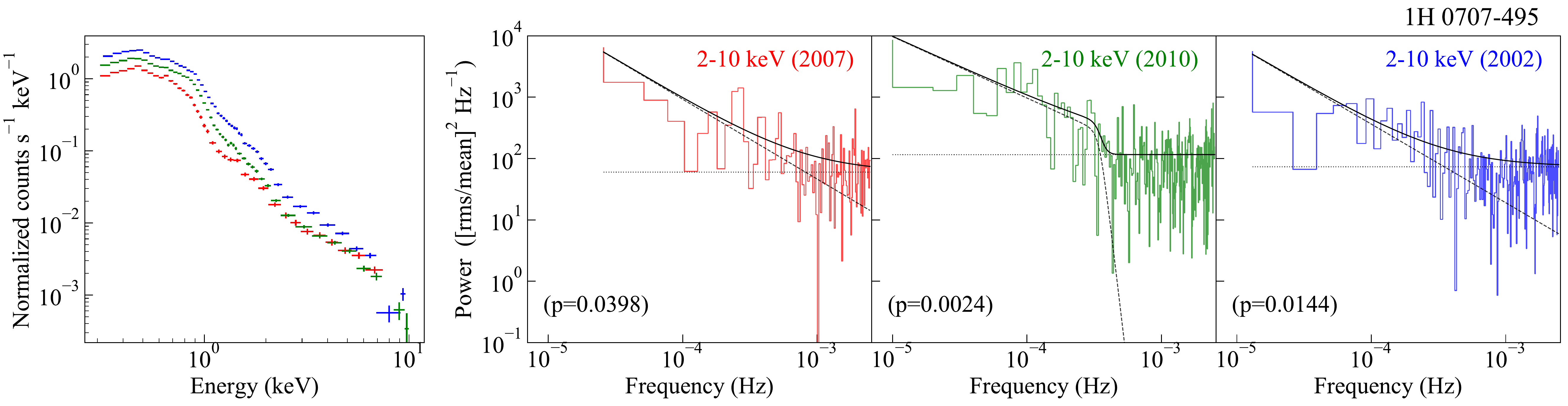}{1.0\textwidth}{}
			}
			\vspace{-0.3cm}
			\gridline{
				\fig{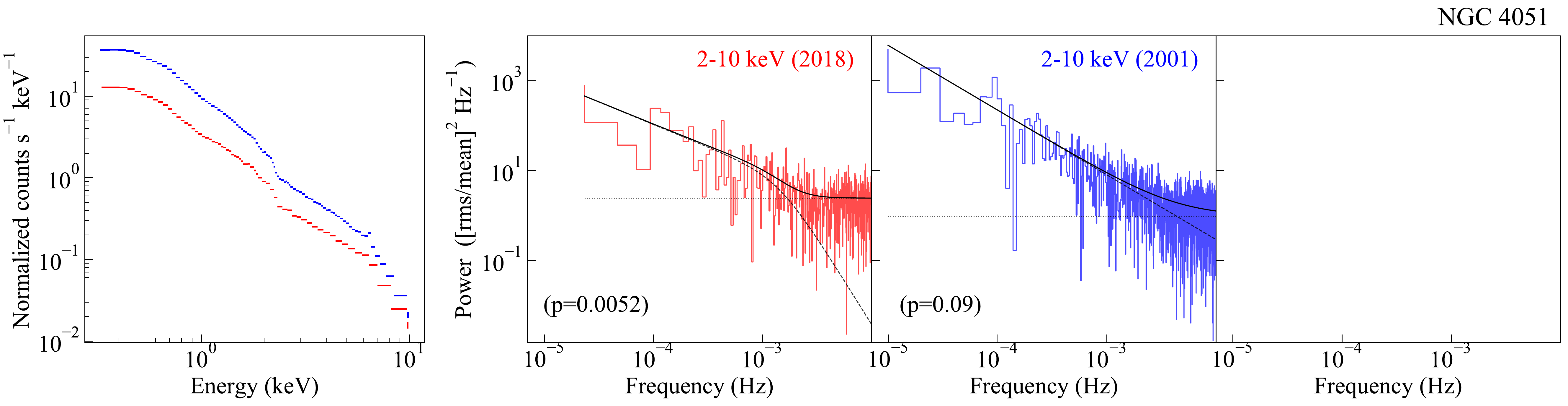}{1.0\textwidth}{}
			}
			\vspace{-0.3cm}
			\caption{Similar to Figure~\ref{fig:spec&PSD_s}, but for the 2-10 keV band. \label{fig:spec&PSD_h}}
		\end{center}
	\end{figure*}

	Based on the methods described above, we investigate the variability of X-ray PSDs and their potential co-variation with the energy spectra. \edit3{{The folded energy spectra are shown in Figures \ref{fig:spec&PSD_s} and \ref{fig:spec&PSD_h}, along with the corresponding PSDs in the soft (0.3-1 keV) and hard (2-10 keV) X-ray bands. And unfolded spectra are plotted in Figure \ref{fig:unfoldspec}.}} For observations with significant breaks in the PSDs the best-fit broken power law model (Model-B) curves are plotted, for the rest the PSDs are shown with the best-fit single power law model (Model-A) curves. The same scale of x-axis and y-axis is adopted so that the difference between observations can be compared directly.
		
\begin{figure}[t!] 
		\begin{center} 
			\includegraphics[width=0.48\textwidth]{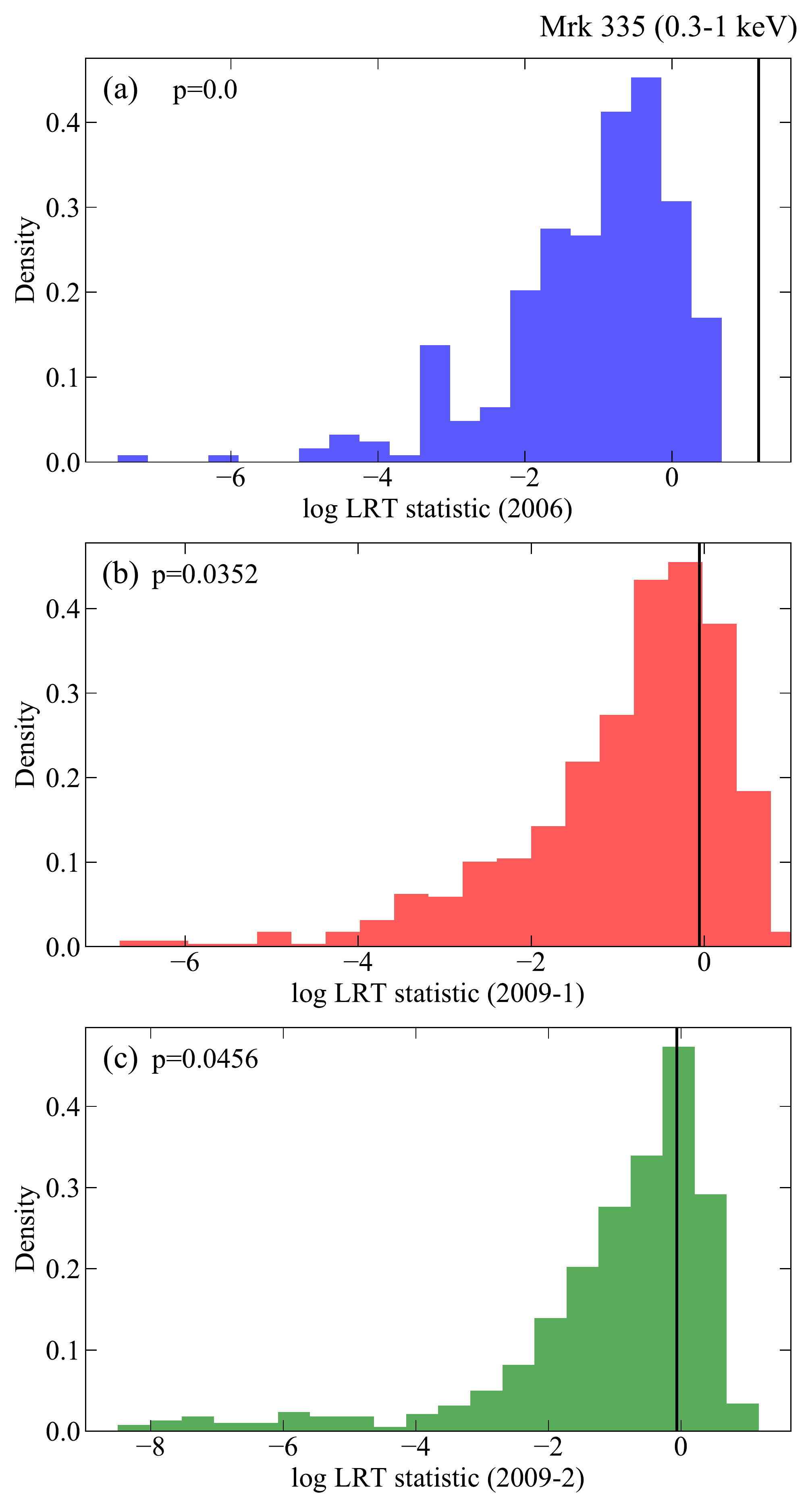}
			\caption{Distributions of LRT1 statistics of Mrk 335 in the 0.3-1 keV band with 5000 datasets simulated under Model-A. The black vertical line is the observed value $T_{\rm LRT}^{\rm obs}$, with the corresponding \textit{p}-value at the top left. \label{fig:mrk335_LRTs_erect}}	
		\end{center}
		\vspace{0.4cm}
	\end{figure}

\begin{figure}[t!] 
		\begin{center} 
			\includegraphics[width=0.48\textwidth]{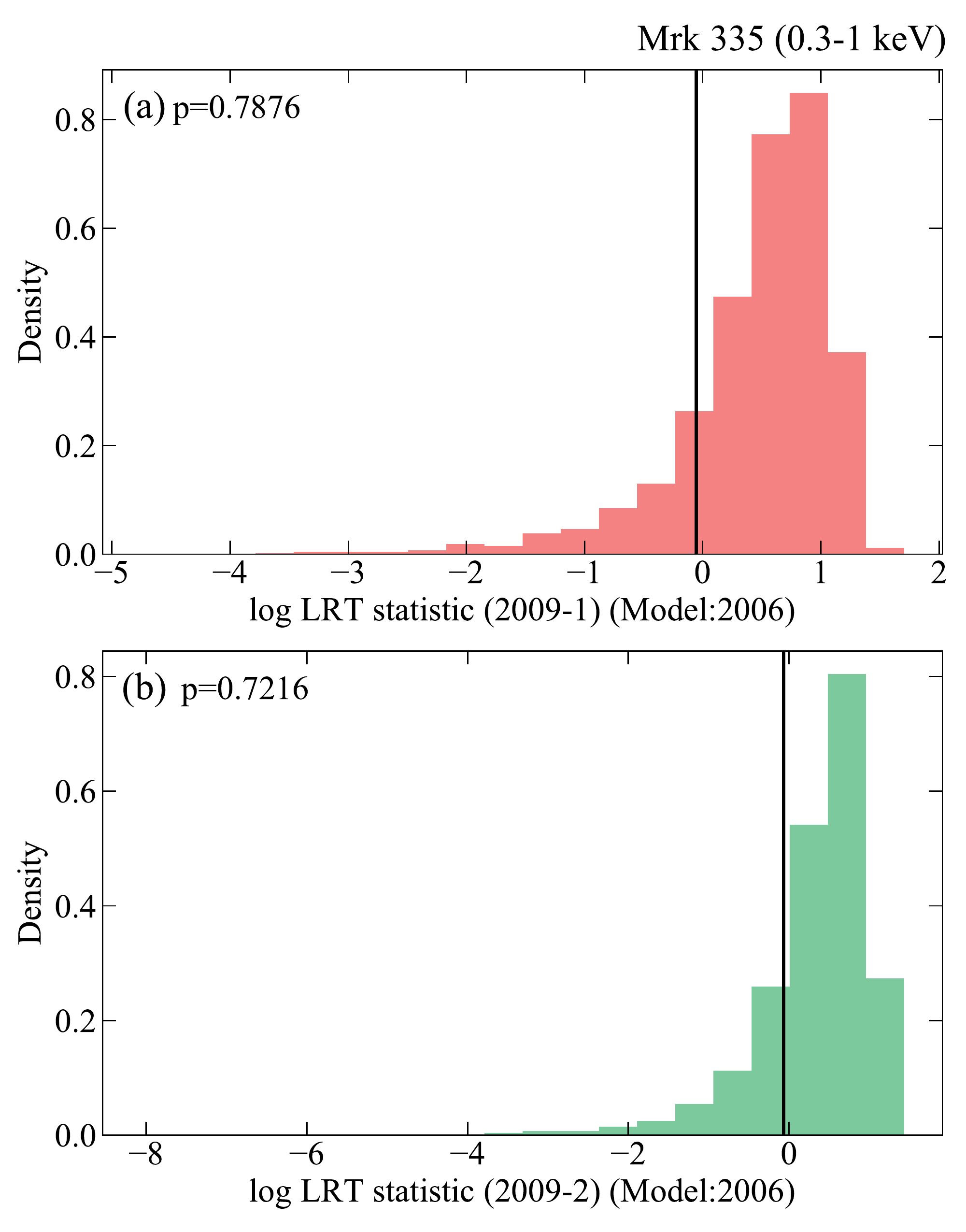}
			\caption{Distributions of LRT2 statistics of Mrk 335 in the 0.3-1 keV band with 5000 datasets simulated under Model-B. The model parameters are based on the posterior distribution for the 2006 observation, while the Poisson term uses the value of the observations in 2009. The black vertical line is the observed value $T_{\rm LRT}^{\rm obs}$, with the corresponding \textit{p}-value at the top left. \label{fig:mrk335_LRTs_bendtest_erect}}	
		\end{center}
		\vspace{0.4cm}
	\end{figure}

	As described in the previous section, we calculated the LRT1 statistics based on the simulation of the best-fit Model-A from the same observation. The distributions of LRT1 statistics for Mrk 335 in the 0.3-1 keV are shown in Figure \ref{fig:mrk335_LRTs_erect}. The LRT1 \textit{p}-values for all sources are shown in Table \ref{tab:LRT1}. The calculation of LRT2 distribution is based on another set of simulation using the best-fit Model-B from a reference observation. The results of LRT2 simulation for Mrk 335 in the 0.3-1 keV are shown in Figure \ref{fig:mrk335_LRTs_bendtest_erect}. The LRT2 \textit{p}-values for all sources are shown in Table \ref{tab:LRT2}.

	\subsection{Results for Individual Sources}
	\subsubsection{Mrk 335}
	The first row of panels in Figure \ref{fig:spec&PSD_s} shows the results for Mrk 335. The 0.3-1 keV periodogram of Mrk 335 from the observation of 2006 shows a significant break, which does not appear in the observations of 2009. This is confirmed by the distribution of LRT1 statistics shown in Figure \ref{fig:mrk335_LRTs_erect}. Among the three observations investigated, only the \textit{p}-value of the observation of 2006 is lower than 0.01. This means that the high-frequency break is significant in the observation of 2006, but not in the two observations of 2009. 
	
	However,  Figure \ref{fig:spec&PSD_s} also shows that the Poisson noise power in the two observations of 2009 is much stronger, and so it dominates the power at lower frequencies, close to the break frequency observed in 2006. Hence the Poisson noise power can reduce the significance of the PSD break signal or even overwhelm it. To overcome this issue, we use the observation in 2006 as the reference observation, and apply the Bayesian MCMC method as described in Section~\ref{sec-psd-compare} to compare these periodograms. As shown by Figure~\ref{fig:mrk335_LRTs_bendtest_erect}, the LRT2 distributions are systematically larger than the observed LRT2. The corresponding \textit{p}-value is 0.79 and 0.72 for the two observations in 2009 (see Table~\ref{tab:LRT2}), which are both much larger than the \textit{p}-value based on Model-A in Table~\ref{tab:LRT1}. This clearly indicates that the difference between the periodograms observed in 2009 and 2006 reflects the intrinsic difference of the PSD, rather than due to the difference of Poisson noise power or the random fluctuation.
		
	Meanwhile, we can see from Figure \ref{fig:spec&PSD_s} that the source flux in the observation of 2006 is much higher than the other two observations, and the spectral shape is also much steeper. The differences between the spectral flux and slope can also be found in Table \ref{tab:obs} quantitatively. Therefore, it seems that for Mrk 335 a high-flux and steeper energy spectrum is associated with the emergence of a high-frequency break in the PSD.
	
	Similarly, as shown in the first panel of Figure \ref{fig:spec&PSD_h}, the 2-10 keV periodogram of Mrk 335 in 2006 is also different from that in 2009, which shows a significant high-frequency break similar to the results found for 0.3-1 keV. The result is robust, as being verified by the same Bayesian MCMC method (see Table~\ref{tab:LRT2}). Besides, the break frequency of 2-10 keV band appears to be higher than that of 0.3-1 keV band, although the difference is not sufficiently significant if the PSD random fluctuation is considered.
	
	It is also interesting to note is that the PSD normalization in the observation of 2006 is lower than that in the two observations of 2009. This suggests that as the flux increases, the relative variability amplitude of Mrk 335 decreases.

	\subsubsection{Mrk 766}
	The PSDs of Mrk 766 for the 0.3-1 keV and 2-10 keV bands are shown in the second row of panels in Figure \ref{fig:spec&PSD_s} and \ref{fig:spec&PSD_h}, respectively, with energy spectra on the left side. The spectrum in 2001, which has the highest flux and steepest slope among the three observations, is associated with the presence of a significant PSD break in the two energy bands.
	
	The other two observations, both of which were made in 2005, are significantly different from each other in terms of their energy spectra and PSDs. The first observation in 2005 does not show high-frequency break in either 0.3-1 keV or 2-10 keV when the spectral flux is low and has a hard shape, which is similar to the observations of Mrk 335 in 2009. The second observation in 2005 shows a relatively higher flux and steeper spectrum, accompanied by the emergence of a PSD break in both energy bands. We have also applied the same method to ensure that the non-detectability of the PSD break is not due to the influence of higher Poisson noise power (see Table~\ref{tab:LRT2}). Interestingly, these results are all similar to those found in Mrk 335.
	
	Likewise, for the two observations which show a high-frequency break, the break frequencies of 2-10 keV are higher than those of 0.3-1 keV, which can be seen in Table \ref{tab:parameters}. The break frequencies detected in the two observations are also statistically consistent with each other. The PSD normalization of Mrk 766 also decreases as the spectral flux increases. Therefore, regarding the flux state, the emergence of PSD break and the PSD normalization, the results of Mrk 766 are the same as Mrk 335.

	\subsubsection{1H 0707-495}
	As in the third row of panels in Figure \ref{fig:spec&PSD_s} and \ref{fig:spec&PSD_h}, the spectral slopes and fluxes of 1H 0707-495 during the observations that we selected have relatively smaller differences than the other sources, but different PSDs are still observed. The PSD in 2007 does not show a high-frequency break in the two energy bands when the energy spectrum has the hardest shape and lowest flux. In comparison, the observations in 2002 and 2010 show high-frequency breaks in the 0.3-1 keV PSDs, and the two break frequencies are consistent with each other. Moreover, only the observation in 2010 show a significant high-frequency break in 2-10 keV. The break does not emerge in the 2-10 keV PSD in 2002, while the spectral flux and slope is highest among the three observations. Likewise, the non-detectability of the PSD break has been checked against the influence of different Poisson noise power (see Table~\ref{tab:LRT2}).
	
	The PSD normalization of 1H 0707-495 is also found to be anti-correlated with the spectral flux. Therefore, for 1H 0707-495 the results in the 0.3-1 keV band are the same as Mrk 335 and Mrk 766. The results in the 2-10 keV band is a little different. A possible explanation is that the break in 2-10 keV in the observation of 2010 is caused by the random fluctuation of the underlying noise.

	\subsubsection{NGC 4051}
	For NGC 4051, only two observations met the requirements that we set in Section~\ref{sec:2}. The spectra and PSDs are shown in the bottom row of panels in Figure \ref{fig:spec&PSD_s} and \ref{fig:spec&PSD_h}. The spectrum in 2001 is steeper than that in 2010, and the flux in 2001 is also higher. But in the 0.3-1 keV band, the break frequency in 2001 is much lower than that in 2018, which is opposite to the behavior of break frequency in other objects. Meanwhile, the 2-10 keV PSD does not show a high-frequency break in the observation of 2001, which has also been confirmed by taking into account the influence of Poisson noise power (see Table~\ref{tab:LRT2}).
	
	Interestingly, the PSD normalization in the observation of 2001 is higher than 2018, except the lower Poisson noise in 2001 due to the higher count rate. This suggests that as the flux of NGC 4051 increases, its relative variability amplitude also increases. Therefore, the link between the PSD normalization and the energy spectrum of NGC 4051 is different from the previous three sources.

	\subsubsection{IRAS 13224-3809, NGC 1365, NGC 3516, \\NGC 4395 and PG 1211+143}
	The other five sources also show highly variable energy spectra, but do not have more than two PSDs which show significant high-frequency breaks. Their energy spectra and PSDs are shown in Figure \ref{fig:spec&PSD_unused_s} and \ref{fig:spec&PSD_unused_h}, while the LRT1 \textit{p}-values used to distinguish PSD breaks are given in Table \ref{tab:LRT1}.
	
	We find four PSD breaks in these observations. They are in the 2-10 keV band in the observation of 2016 for IRAS 13224-3809, the 2-10 keV band in the observation of 2013 for NGC 1365, the 2-10 keV band in the observation of 2019 for NGC 4395, and the 0.3-1 keV band in the observation of 2001 for PG 1211+143. The LRT1 \textit{p}-value are lower than 0.01 for these four PSDs. However, for these datasets, the MCMC runs did not converge except the 2019 observation of NGC 4395. This result suggests that these PSD breaks are poorly constrained. Although the break in these observations can exist, it is elusive to compare these PSD shapes in more detail for the lack of other break detection. Moreover, for NGC 3516 and NGC 4395 we notice that as the spectral flux decreases, the intrinsic variability decreases as well.

	\subsection{Overall Results for the Sample}	
	
	\edit1{{
	All our sources are included in the sample of \citet{Gonzalez-Martin2012}, which is also the parent sample of \citet{Kara2016b}. Our fitting results are generally consistent with those in \citet{Gonzalez-Martin2012}. However, our analyses go further given that we focus on PSD changing between observations of each source. For example, two observations close in time were analyzed together in \citet{Gonzalez-Martin2012}, while we analyzed them separately and pay attention to the variation of PSD if the spectrum changes significantly (e.g. two observations of Mrk 766 in 2005). The break frequencies found in this work are generally consistent with those reported by \citet{Gonzalez-Martin2012}, although some differences do exist. Specifically, our results are slightly higher for Mrk 766 in 2001, Mrk 766 in 2005, and 1H 0707-495 in 2010. For NGC 4051 in 2001, \citet{Gonzalez-Martin2012} report a PSD break in 2-10 keV at $\log(f_{\mathrm{B}})\sim -3.99$ with a large uncertainty, but this is not detected by our analysis. These differences are likely caused by different methods of PSD analysis. The observation of NGC 4051 in 2018 is not included in \citet{Gonzalez-Martin2012}, yet studies about this source has shown that it has unique spectral variability which is more than ``softer-when-brighter'' trend and can be explained by a geometry include an extended corona \citep{Wu2020}. Timing analysis of these sources may provide more clues in addition to the spectral property. 
	}}
	
\begin{figure*}[t!] 
		\begin{center} 
			\gridline{\fig{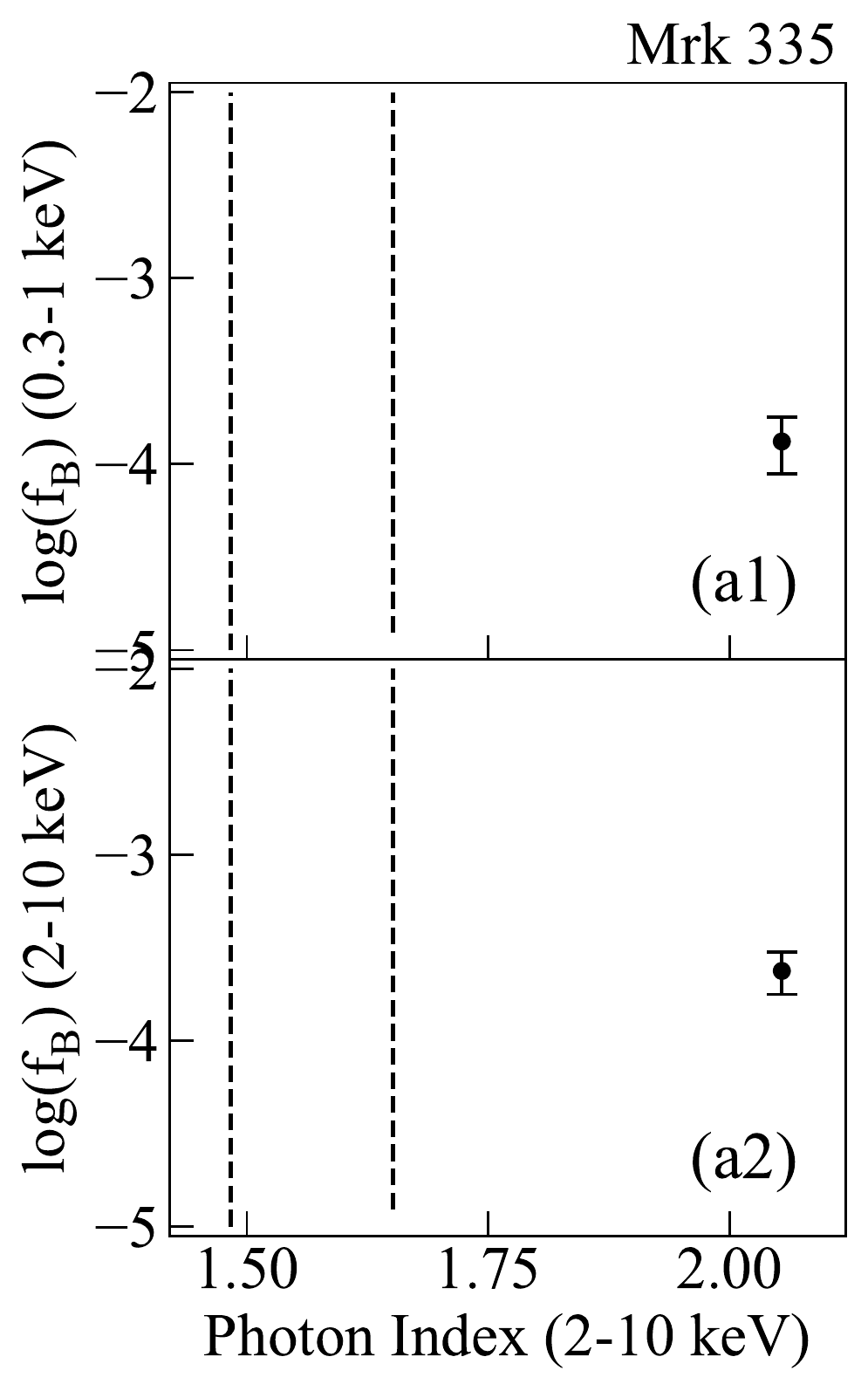}{0.23\textwidth}{}\fig{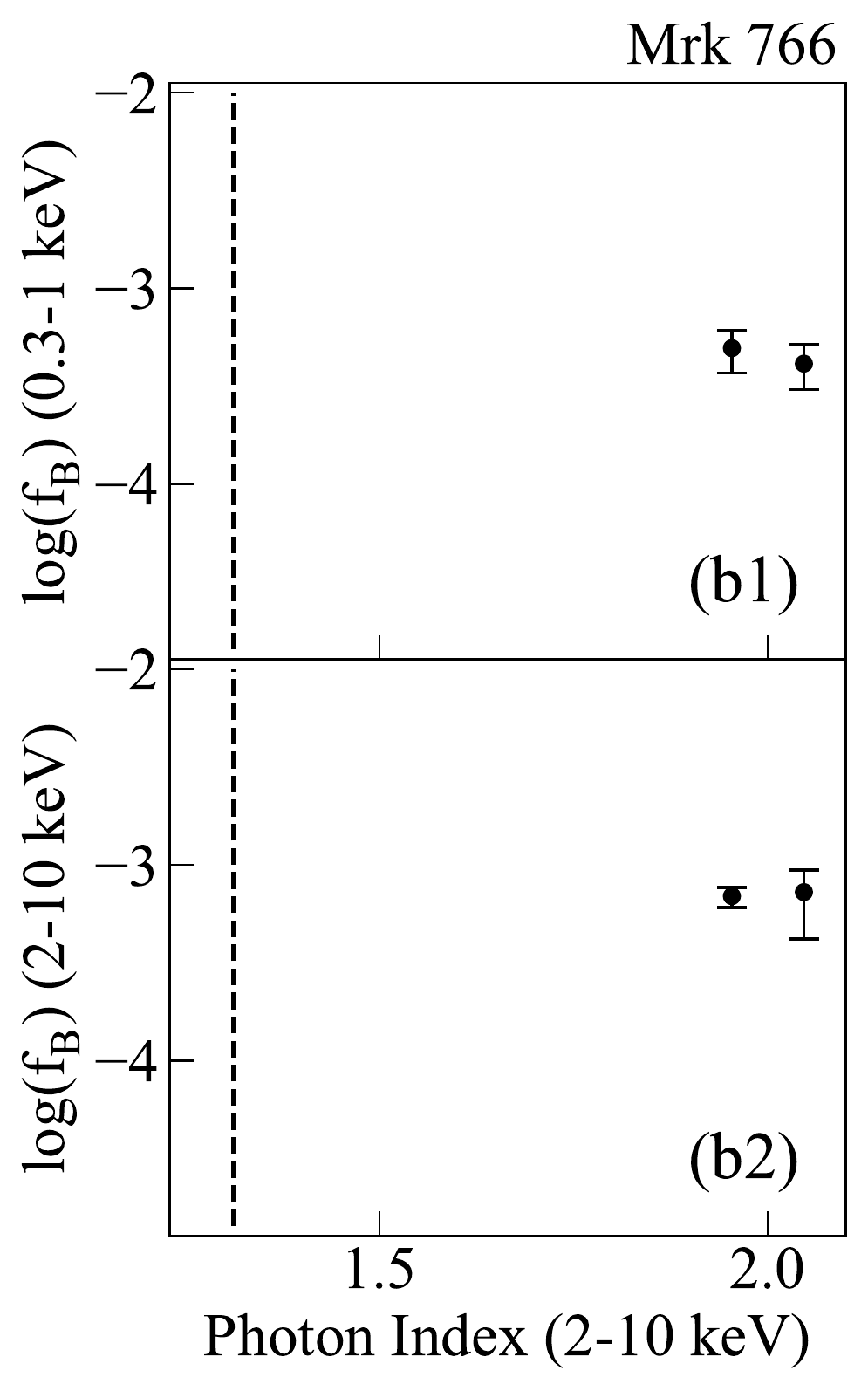}{0.23\textwidth}{}\fig{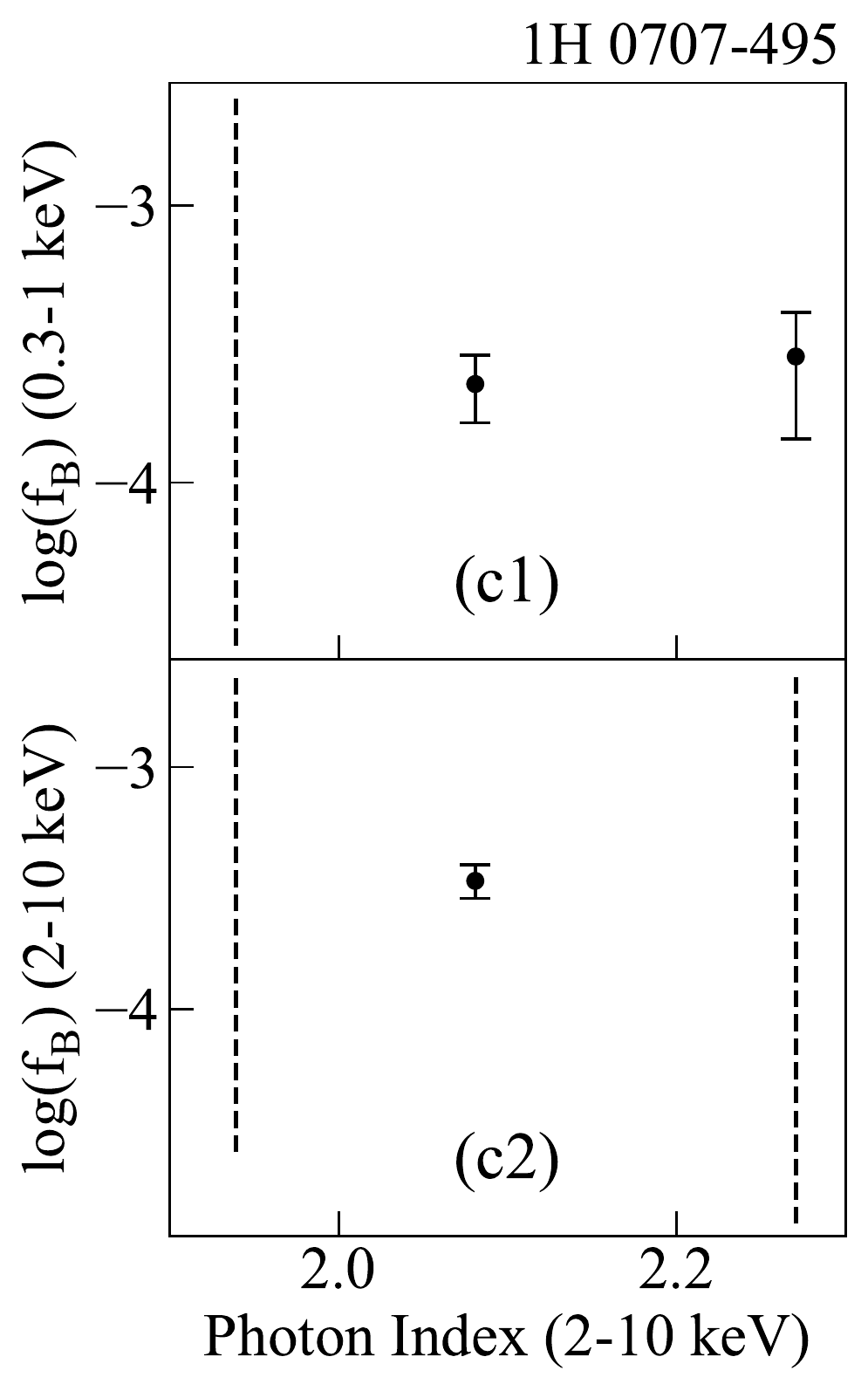}{0.23\textwidth}{}\fig{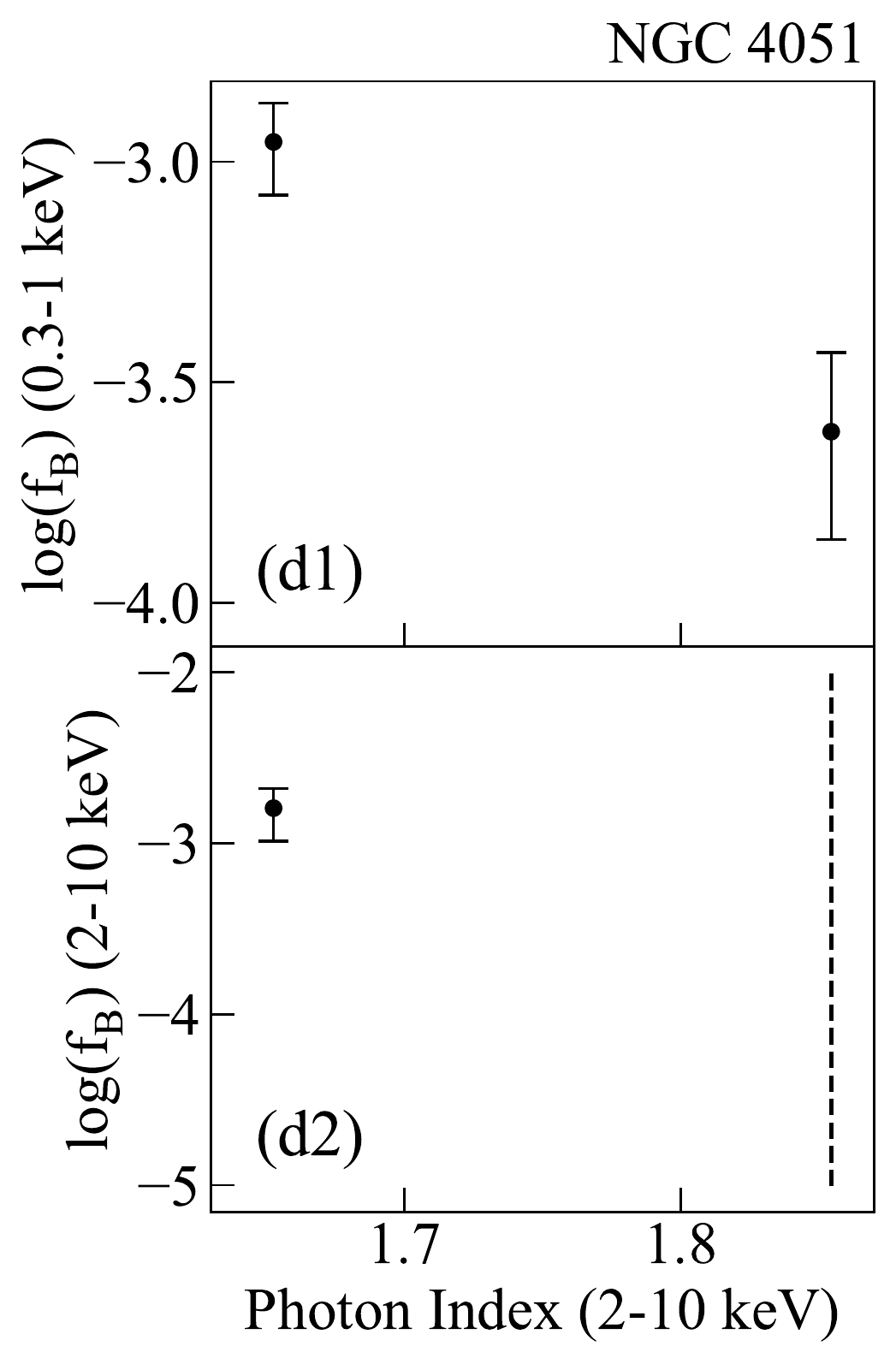}{0.242\textwidth}{}}
			\vspace{-0.6cm}
			\caption{Logarithm of the PSD break frequencies against spectral slopes. The values and errors are calculated using MCMC. The data are marked as dashed lines which show the observed frequency ranges of PSDs if they do not exhibit a significant break. \label{fig:logf_slope}}
		\end{center}
	\end{figure*}

\begin{figure*}[t!] 
		\begin{center} 
			\gridline{\fig{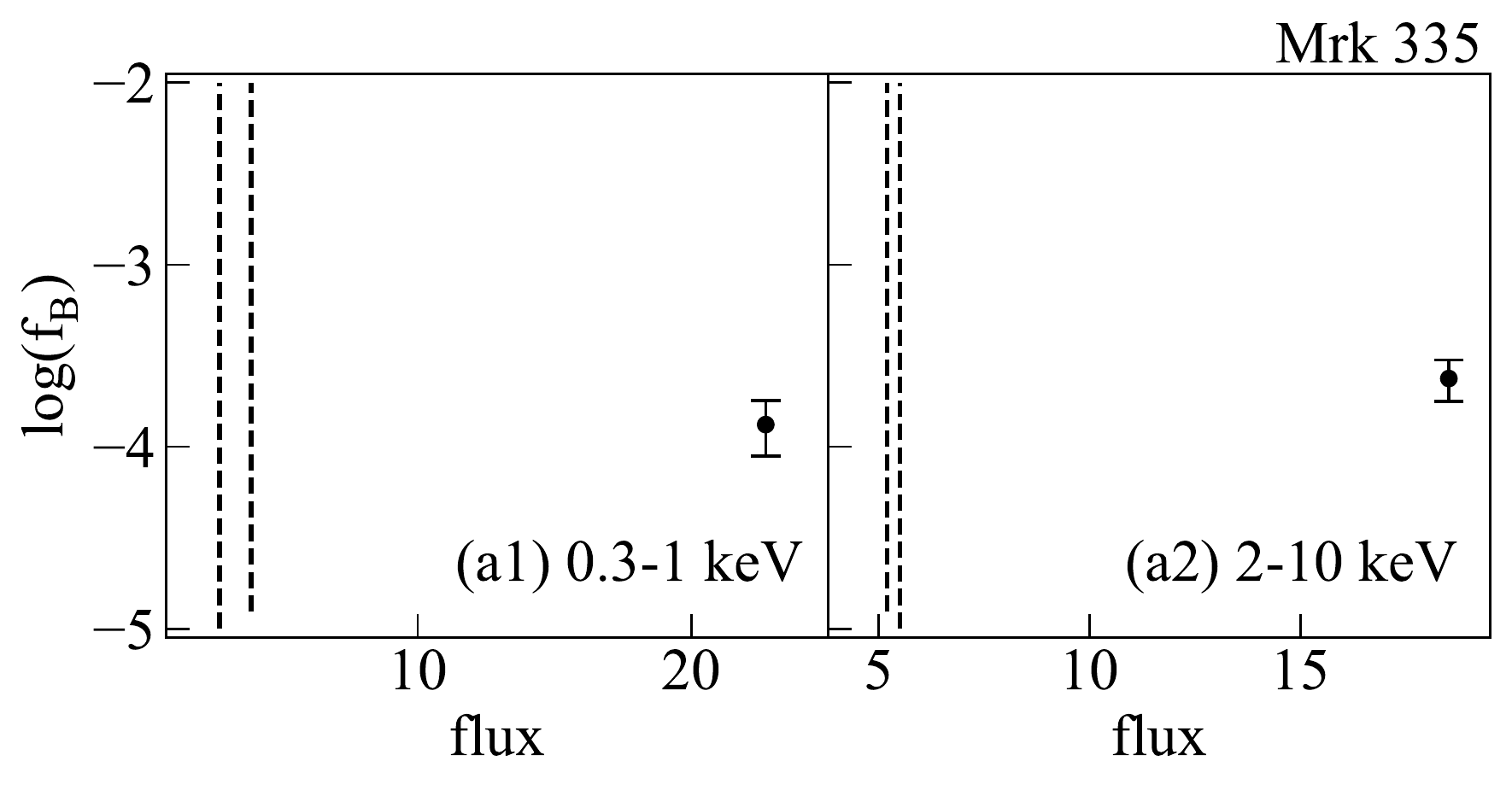}{0.46\textwidth}{}
				\fig{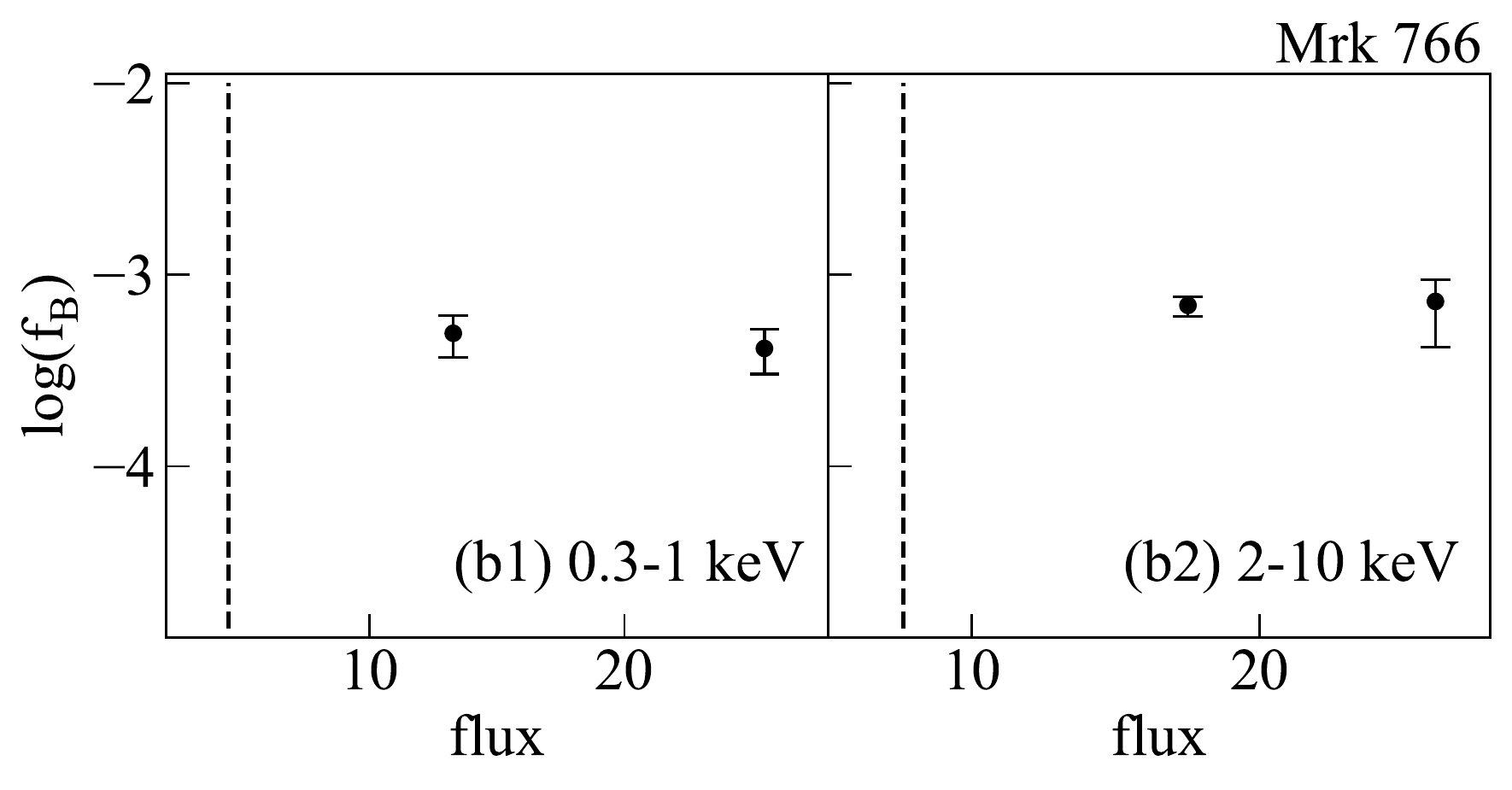}{0.46\textwidth}{}
			}
			\vspace{-0.6cm}
			\gridline{\fig{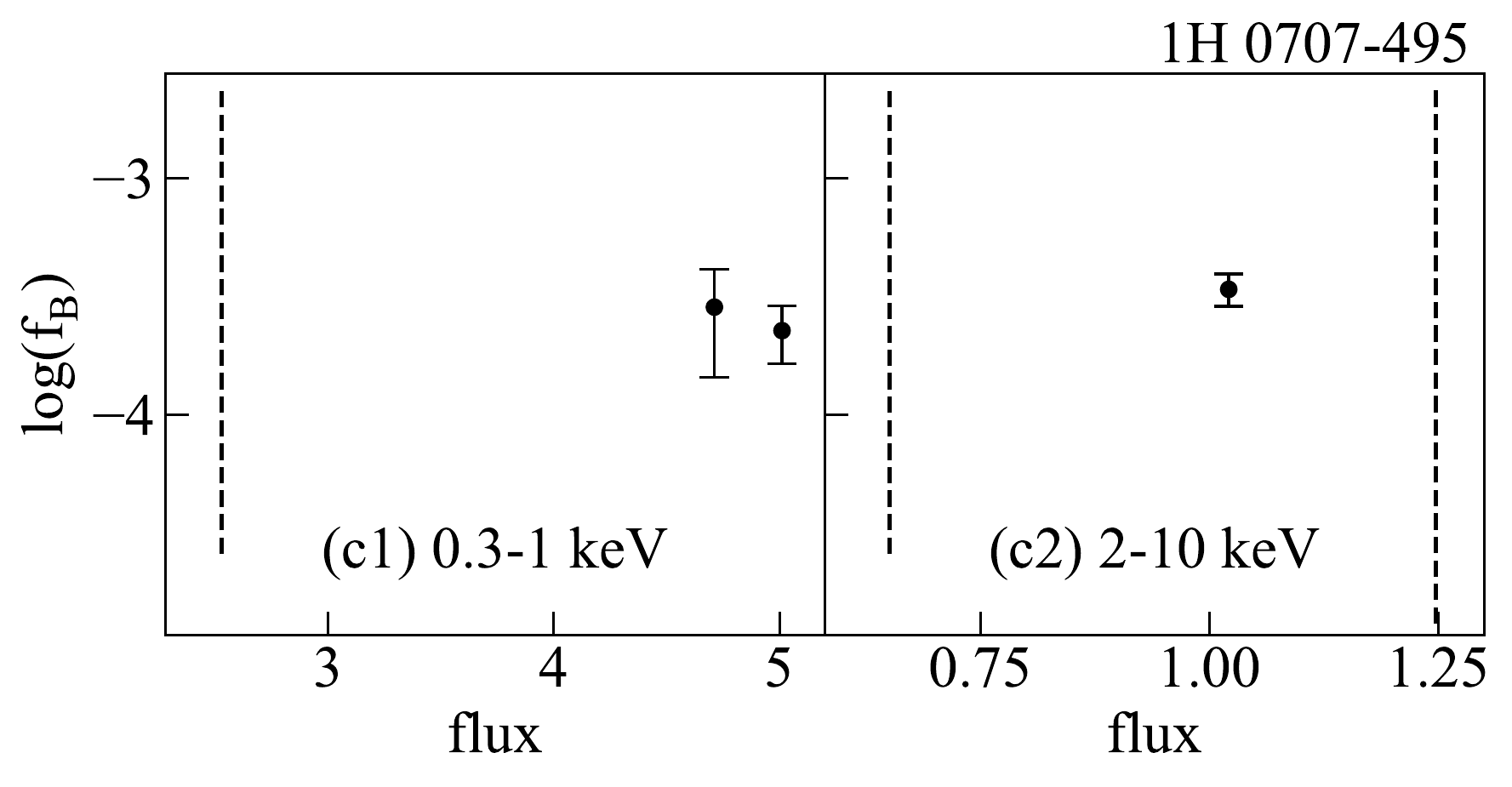}{0.46\textwidth}{}
				\fig{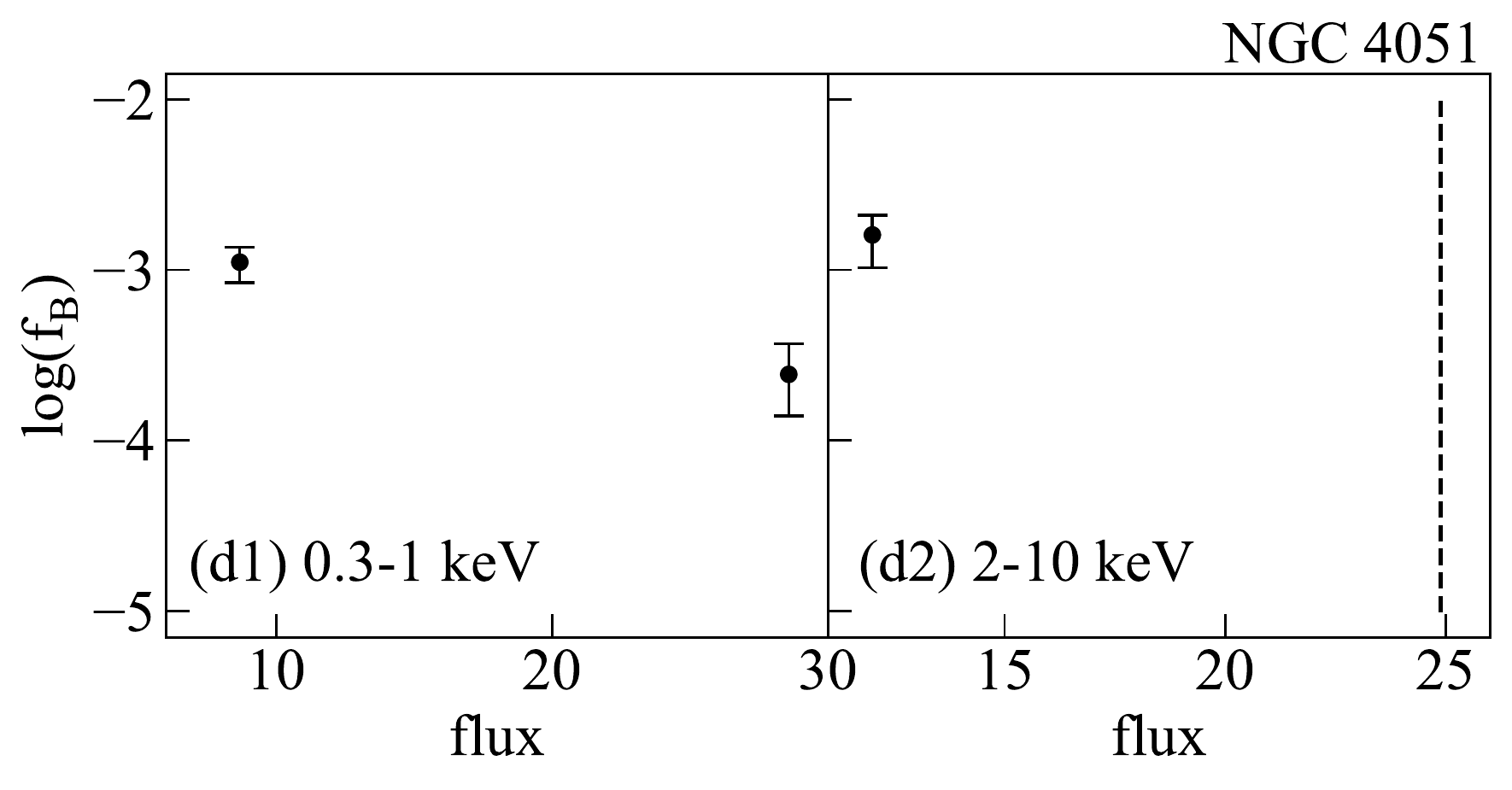}{0.46\textwidth}{}
			}
			\vspace{-0.6cm}
			\caption{Logarithm of the PSD break frequencies against fluxes. The values and errors are calculated using MCMC. The data are marked as dashed lines which show the observed frequency ranges of PSDs if they do not exhibit a significant break. \label{fig:logf_flux}}
		\end{center}
		\vspace{0.4cm} 
	\end{figure*}

	We compare the relations between the PSD break frequencies and the spectral slopes and fluxes for Mrk 335, Mrk 766, 1H 0707-495 and NGC 4051 in Figures \ref{fig:logf_slope} and \ref{fig:logf_flux}. These AGN showed significant PSD breaks in at least two \textit{XMM-Newton} observations. For observations in which no PSD break is detected, we use a dashed line to indicate the observed frequency range of the PSD in which there is no detectable break.
	
	As we pointed out before, for Mrk 335, Mrk 766, 1H 0707-495, the PSD break is detectable only when the spectral flux is high and the shape is steep (the photon index is $\gtrsim$ 2.0). A possible explanation is that their low-flux spectra are dominated by additional processes. Indeed, these sources all have very high mass accretion rates, and it has been proposed that their drastic spectral variability may be caused by the absorption of strong disc wind (e.g. \citealt{Done2016a}; \citealt{Hagino2016}). Hence a natural explanation would be the absorption of wind clumps, which may introduce extra variability and lead to the undetectability of the PSD break during the low-flux state.
	
	However, the situation of NGC 4051 is somewhat different. Both the high and low-flux spectral states show PSDs with a high-frequency break, and the break frequency is higher when the flux is lower. Therefore, we conclude that in a low-flux state, the PSD shows either no break or a break shifted to a higher frequency where it is hard to detect the break for the PSD at high frequency is dominated by white noise. Physically, the main difference of NGC 4051 is that its mass accretion rate is much lower than those of the other three sources, hence the origin of its spectral variability is probably different.
	
	Then we investigate if there is any correlation between the variations of spectra and  PSDs for all the four sources. As shown in Figure \ref{fig:spec&PSD_s} and \ref{fig:spec&PSD_h}, for observations showing drastically different states in flux or spectral shape, their PSDs also show significant differences. This can be confirmed by the changing of LRT1 \textit{p}-values of observations from the same source, and difference between LRT1 \textit{p}-value and LRT2 \textit{p}-value using the same observed $T_{\rm LRT}^{\rm obs}$ but with two simulation sets. Therefore, our result hints at a possible correlation between the spectral state and PSD, although the sample size is too small to draw a general conclusions for AGN.

\section{Discussion} \label{sec:5}
	\subsection{Variation of the X-ray PSD in AGN}
	It is well known that there is a coevolution between the spectral state and the PSD shape for XRBs. However, the variability of AGN's PSD is not well explored. In this work, we attempt to look for evidence of more general correlations between AGN's energy spectra and PSDs. Sources in our sample all exhibit drastic spectral variation, which increases the possibility to observe the change of PSD caused by the change of underlying physical processes rather than random fluctuation.
	
	Indeed, we find that the PSD changes in both the shape (the detectability and frequency of the high-frequency break) and normalization (i.e. the rms amplitude).
	\edit1{{In our sample, observations of individual sources with similar energy spectra tend to show similar PSDs, while a large difference in the spectral shape is also accompanied by a large difference in the corresponding PSD shape.}} 
	This indicates that in AGN there is also a tentative link between the X-ray spectral state and PSD. \edit1{{However, this potential correlation is limited by the sample size. A larger sample is needed to verify it.}}
	
	Moreover, we find that for Mrk 335, Mrk 766 and 1H 0707-495, the high-frequency break only appears in the high-flux state \edit1{{in the observed frequency range}}. For NGC 4051, the break frequency in the low-flux state is higher than in the high-flux state. These results are different from the coevolution of energy spectrum and PSD observed in XRBs. Therefore, it is likely that the high and low-flux spectral states observed in our AGN sample do not correspond to different accretion states as in XRBs. Indeed, previous studies of 1H 0707-495 have shown that its X-ray flux can change by more than one order of magnitude while no significant variation of accretion rate is observed in the outer disc (e.g. \citealt{Done2016a}). This is also consistent with the expectation that the timescale of accretion state transition is much longer than the variability timescales covered by individual \textit{XMM-Newton} observations. Therefore, the driver behind the correlation between the PSD and the energy spectrum is not likely to be the accretion rate. Another possible scenario for the observed correlation is that the changes in the energy spectrum are mainly caused by the absorption of interfering materials along the line of sight (e.g. \citealt{Connolly2014}), and/or reflection from disc wind or outflowing materials (e.g. \citealt{Miller2010}; \citealt{Done2016a}; \citealt{Hagino2016}). The low-flux state suffers more severe absorption, which suppresses the intrinsic X-ray variability and causes the PSD break to disappear.
	
	The result of NGC 4051 is also consistent with the positive correlation between $N_{\rm H}$ and $f_{\rm B}$ found by \citet{Gonzalez-Martin2018}, in which the harder the spectral shape is and the lower the spectral flux is, the higher the break frequency would be. A simple scenario is that the X-ray corona has an extended structure, where soft X-rays come from larger radii than hard X-rays, thus their corresponding variability timescales are longer (e.g. \citealt{Jin2013, Jin2017}). Therefore, as more soft X-rays are absorbed, the observed energy spectrum becomes harder, and the observed variability timescale becomes shorter. This hypothesis, which can be supported by the fact that the break frequency in the hard X-ray band is higher than in the soft X-ray band, should be tested with more careful studies including detailed spectral fitting and variability analysis. The additional variability caused by the reflection of the absorbing material may also provide an explanation (\citealt{Miller2010}), but this is also related to the disc wind or outflowing materials whose properties and distributions may be different from one source to anther. More detailed spectral analysis for each source is required, but it is beyond the scope of this work.

	\edit1{{We also note that the Eddington ratios of most of our sources are higher than a few per cent (see Table \ref{tab:mass_of_sources}), and so their accretion flows should behave like a standard sub-Eddington thin disc (e.g. \citealt{Shakura1973}) or a super-Eddington disc (e.g. \citealt{Poutanen2007}). Only NGC 4395 shows an Eddington ratio lower than 0.01. However, given that the uncertainty of black hole mass estimation is large, the calculated Eddington ratio could not be accurate. Hence the possibility that the accretion rate of NGC 4395 is higher than estimation and a normal disc exists can not be ruled out. In comparison, XRBs can also evolve to a low/hard state when their Eddington ratios drop below $\sim~0.02$ (e.g. \citealt{Maccarone2003}) and becomes radiative inefficient (e.g. \citealt{Narayan1995}). Therefore, our AGN sample cannot map to every type of accretion state of XRBs.}}

	\edit1{{For XRBs the frequency range used to study the PSD evolution with other parameters is 4-5 decades (e.g. \citealt{Gierlinski2008}). It was found that all the characteristic frequencies increases as the mass accretion rate increases (e.g. \citealt{vanderKlis2004, Axelsson2005, Remillard2006}), which can be understood in the context of truncated disc model (e.g. \citealt{Done2003, Done2007}). In this work, the frequency range of AGN PSD available for comparison is only 2 decades, i.e. $10^{-5}-10^{-3}$Hz, before Poisson noise dominates higher frequencies. Hence, it is possible that the break frequency of the observations without break detection is located outside the frequency window. The fitting results of Model A shown in Table \ref{tab:parameters_modelA} with slope $\alpha$ significantly larger than 1 may provide some implication on the existence of such a break at lower frequencies. Indeed, the 2-10 keV PSD of NGC 3516 was reported to have a high-frequency break at $f_\mathrm{B}\sim 2.00\times 10^{-6} Hz$ by \citet{Markowitz2003}, although the data sets are also possibly be affected by the red-noise leak \citep{Gonzalez-Martin2012}.
	}}
	
	\edit1{{
	The motivation for the PSD fitting of Model B by a broken power law model from slope of -1 (i.e. $\alpha_{\mathrm{L}}=1$) to slope steeper than -2 (i.e. $\alpha_{\mathrm{H}}\textgreater 2$) comes from PSDs of XRBs in the soft states and has been tested in many long-term studies of AGN (e.g. \citealt{Uttley2002, Markowitz2003, McHardy06}). Although the choice of $\alpha_{\mathrm{L}}$ for Model B can change the fit result of break frequency, which has been tested in some sources of our sample, a fixed slope at lower frequency is more reasonable considering the lack of power spectral data points and the window effect from time bin division at lower frequency.
	}}
	
	\subsection{Estimating Black Hole Mass Using X-ray Variability}
	It is generally believed that the hard X-ray corona of AGN is located near black hole's event horizon, and so the properties of the corona can be used to infer the properties of the black hole. Indeed, it has been found that the black hole mass ($M_{\rm BH}$) is significantly correlated with the hard X-ray excess variance ($\sigma^2_{\rm rms}$) \citep{Zhou2010, Ponti2012, Pan2015}. It is also known that the black hole mass is anti-correlated with the break frequency ($f_{\rm B}$) of the X-ray PSD \citep{McHardy06, Gonzalez-Martin2012, Gonzalez-Martin2018}. These two correlations are important as they are widely used to obtain estimates of the black hole mass independent from optical estimators (e.g. the virial mass based on the gas dynamics of the broad line region: \citealt{Peterson2004}).
	
	\edit1{{
	The normalized excess variance is calculated following \citet{Nandra1997} using the following equation:
	\begin{equation}
		\sigma^2_{\mathrm{rms}}=\frac{1}{N\mu}\sum_{i=1}^N[(X_i-\mu)^2-\sigma_i^2],
	\end{equation}
	where $N$ is the number of time bins of the light curve, $\mu$ is the unweighted mean count rate, $X_i$ and $\sigma_i$ are the count rates and errors in every time bin.
	Following \citet{Ponti2012}, every light curve is binned in 250 s. The selection criteria of our sample is to have exposure time longer than $40$ ks, and so we can divided light curves into segments of 20 ks length, and calculate the excess variance for every segment. The mean of segments is taken if an observation can be divided into more than one segment. The uncertainty of the excess variance is calculated following \citet{Vaughan2003a},
	\begin{equation}
		\Delta \sigma^2_{\mathrm{rms}}=\sqrt{(\sqrt{\frac{2}{N}}\frac{\langle \sigma^2_i \rangle}{\mu^2})^2 + (\sqrt{\frac{\langle \sigma^2_i \rangle}{N}}\frac{2F_{\mathrm{var}}}{\mu})^2},
	\end{equation} 
	where $\langle \sigma^2_i \rangle$ is the mean of the square of count rate errors, and $F_{\mathrm{var}}=\sqrt{\sigma^2_{\mathrm{rms}}}$ is the fractional variability.
	}}

	However, significant dispersions have been found in these two correlations. For example, \citet{Ponti2012} reported a dispersion of $\sim$ 0.7 dex for the $\sigma^2_{\rm rms}$-$M_{\rm BH}$ relation. Since $\sigma^2_{\rm rms}$ is the integration of the X-ray PSD, the variability of PSD naturally lead to the variability of $\sigma^2_{\rm rms}$ for the same AGN. This is a possible origin for the dispersion observed in \citet{Ponti2012}, which can be tested with our sample.
	
\begin{figure}[t!] 
		\begin{center} 
			\includegraphics[width=0.47\textwidth]{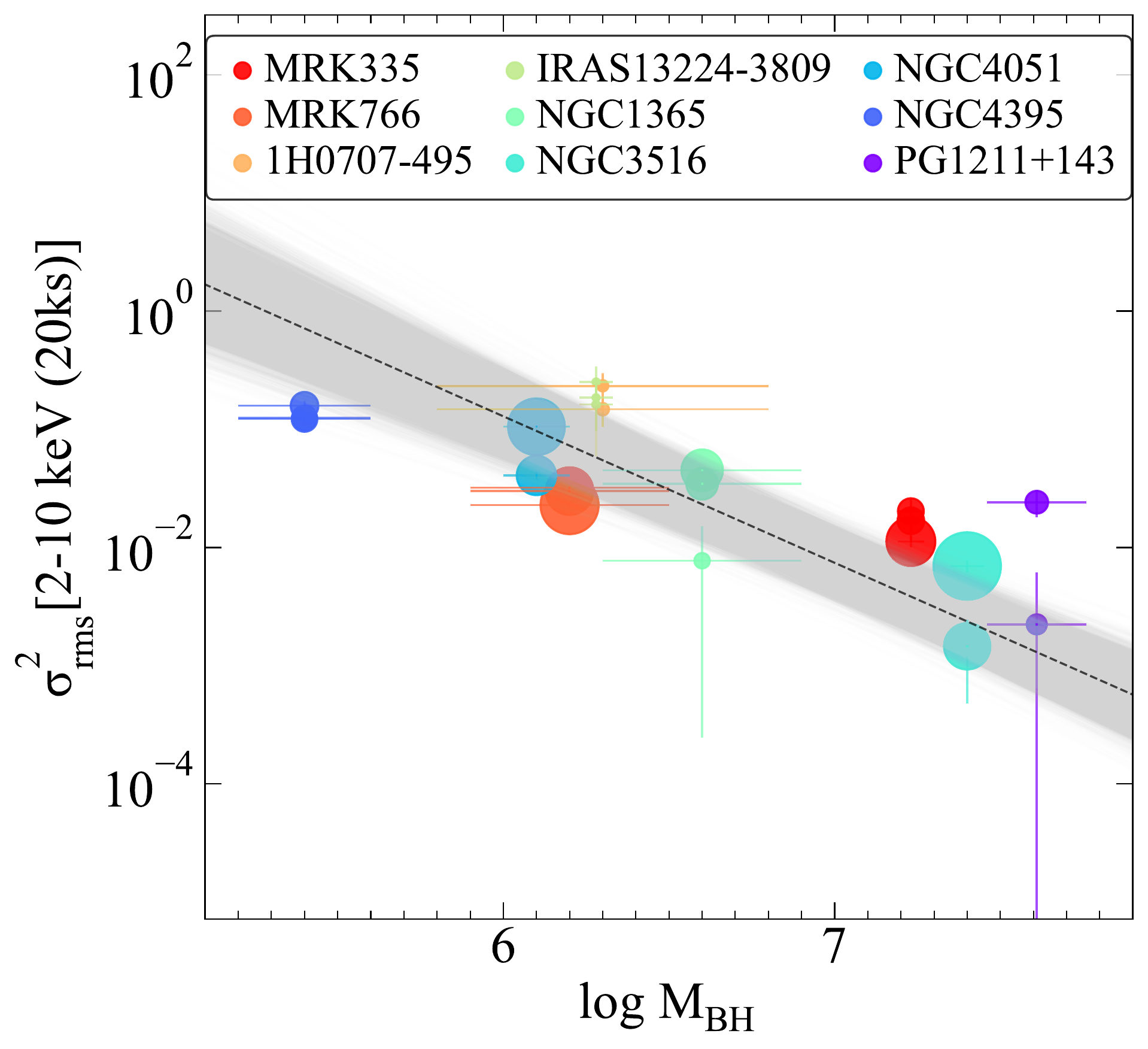}
			\caption{Excess variance calculated using 20 ks segments of light curve against the black hole mass. The sizes of points indicate the relative 2-10 keV flux of corresponding observations. The best fit relationship in \citet{Ponti2012} are marked as the black dotted line \edit1{{and the 1-$\sigma$ error region are shown as a gray shadowed area which comes from random sampling under distributions of slope and normalisation from \citet{Ponti2012}}}.  \label{fig:rms20_h}}
		\end{center}
		\vspace{0.4cm} 
	\end{figure}

	We plot the $\sigma^2_{\rm rms}$ and $M_{\rm BH}$ of our sample in Figure \ref{fig:rms20_h} using the black hole masses \edit1{{and errors}} listed in Table \ref{tab:mass_of_sources}. Our data points are generally consistent with the linear correlation reported by \citet{Ponti2012} for the 20 ks segments, which is shown by the dotted line. Only the points of PG 1211+143 show a relatively large deviation, which is probably caused by the poorly constrained black hole mass of this source \citep{Peterson2004}.
	
	Our sample also shows significant variation in $\sigma^2_{\rm rms}$, with large variation seen in NGC 1365, NGC 3516 and PG 1211+143, which is $\sim$ 1.0 dex, corresponding to $\sim$ 0.8 dex of the black hole mass estimate. Hence, the variation of $\sigma^2_{\rm rms}$ can indeed contribute to the dispersion observed in the $\sigma^2_{\rm rms}$-M$_{\rm BH}$ relation, but it should not be the only source of dispersion, because there must be some dispersion caused by the measurement uncertainty of the black hole mass.
	
\begin{figure}[t!] 
		\begin{center} 
		    \includegraphics[width=0.47\textwidth]{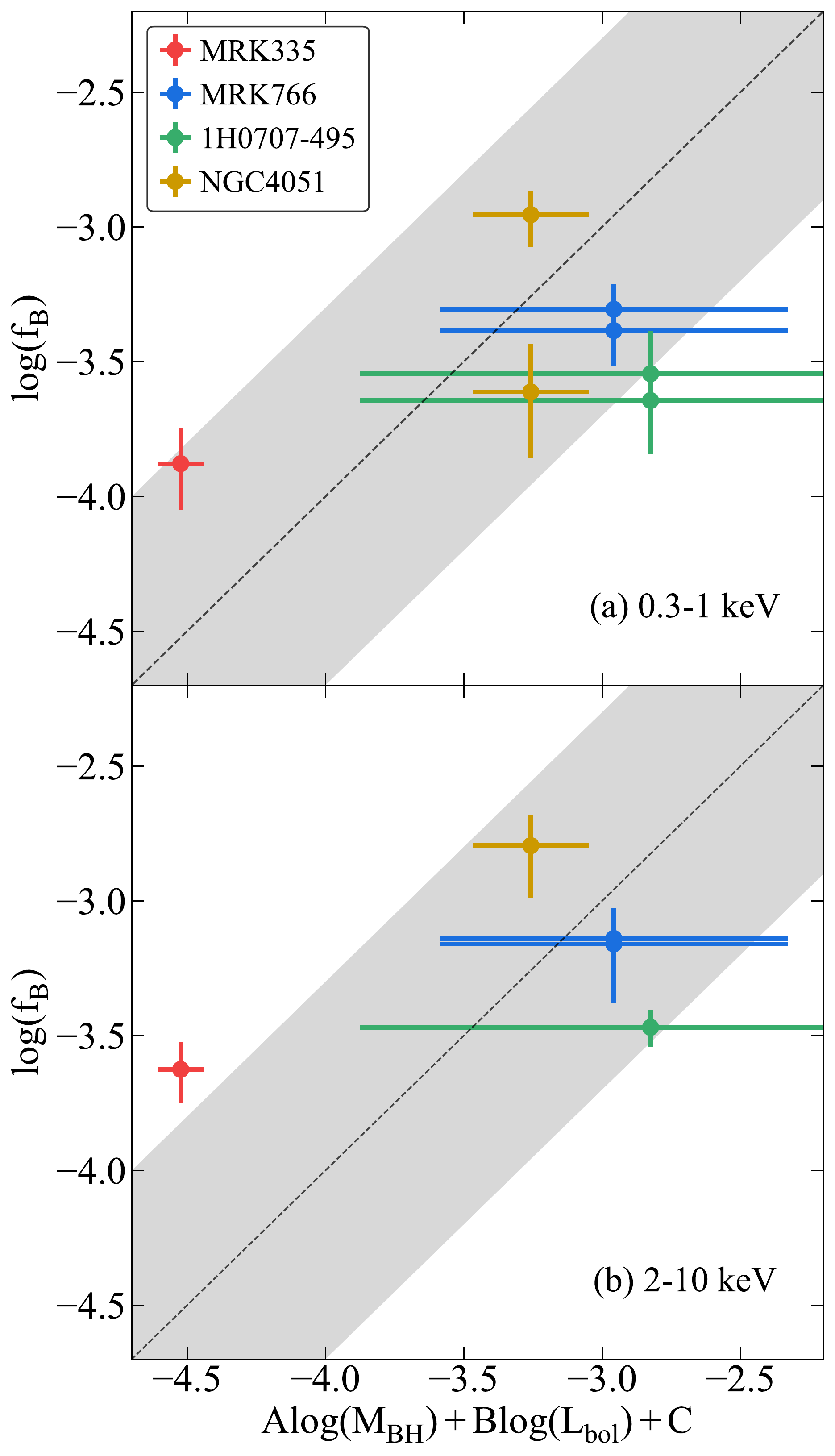}
			\caption{The relationship between the PSD break frequency ($f_{\rm B}$) and black hole mass ($M_{\rm BH}$) and bolometric luminosity ($L_{\rm bol}$). The values of horizontal axis are calculated using expression given in \citet{McHardy06} (which is shown as the diagonal black dotted line), while the observed break frequencies are got in our sample. \edit1{{The uncertainties of mass estimation are used to get the horizontal errors. The gray shadowed areas display $\pm 0.7$ dex regions around the model, as a nearer study of \citet{Gonzalez-Martin2018} show standard deviations around this value.}} \edit1{{Related parameters}} are listed in Tables \ref{tab:parameters} and \ref{tab:mass_of_sources}. \label{fig:plane_mchardy}}
		\end{center}
		\vspace{0.4cm} 
	\end{figure}

	In comparison, the $f_{\rm B}$-$M_{\rm BH}$ relation appears more dispersed. It has been reported that this relation may also depend on the bolometric luminosity ($L_{\rm bol}$, \citealt{McHardy06}) or the neutral absorption column ($N_{\rm H}$, \citealt{Gonzalez-Martin2018}). The latter dependence is such that as the $N_{\rm H}$ increases, $f_{\rm B}$ also increases. In fact, $N_{\rm H}$ in \citet{Gonzalez-Martin2018} is derived from fitting the 2-6 keV spectrum, so it can also be understood as an indicator for the hard X-ray shape. Then a larger $N_{\rm H}$ would indicate a smaller photon index, and then a larger $f_{\rm B}$ is expected. We note that NGC 4051 does follow this trend, but the other sources in our sample do not. We also plot our sample in Figure \ref{fig:plane_mchardy} using the PSD breaks detected in both 0.3-1 keV and 2-10 keV listed in Table \ref{tab:parameters} and parameters listed in Table \ref{tab:mass_of_sources}. \edit1{{Our bolometric luminosities are taken directly from previous works (e.g. \citealt{Woo2002, Zhou2005a, Vasudevan2010, Meyer-Hofmeister2011}). It must be noted that the systematic uncertainty of the bolometric luminosity could be large, as most of the energy of an AGN spectral energy distribution is contained in the far UV band which is not observable (e.g. \citealt{Jin2012}). For the ease of comparison with previous works (e.g. \citealt{McHardy2016}), we did not consider the uncertainty of bolometric luminosity in our analysis. Besides, although the X-ray flux changes in different spectral state, it only contribute a small fraction of the bolometric luminosity (e.g. \citealt{Vasudevan2010, Jin2012}), and so it should not affect our results significantly.}}
	
	As shown by Figure \ref{fig:plane_mchardy}, our data points are roughly consistent with the relation reported by \citet{McHardy06} which is shown as the dotted line. The relatively large variation of break frequency showed in NGC 4051 brings a $\sim$ 0.3 dex uncertainty to the black hole mass estimation, without considering the changes of other parameters. However, \edit1{{although we only fit the spectra with a fixed Galactic absorption $N_{\mathrm{H}}$ given that this value is constant for different observations of the same object,}} few sources of our sample show significant absorption when fitting the 2-6 keV spectrum with an \edit1{{unfixed}} absorbed power law model, which makes it hard to test the relation reported by \citet{Gonzalez-Martin2018}. \edit1{{Besides, although the choice of spectral band from 2-6 keV is more conservative considering the effect of reflection component like Fe line on spectrum and variability, the 2-6 keV light curves and PSDs do not show significant difference from that in 2-10 keV.}}

	\edit1{{In summary, our results suggest that the variation of PSD, which is linked to the spectral state of AGN, can provide a significant fraction of the uncertainty of these scaling relations, so it is recommended to consider the intrinsic variation of the PSD when using AGN X-ray variability parameters (e.g. the rms and the high-frequency break) to estimate the black hole mass.}} In principle, it is possible to take into account spectral variability and to reduce the dispersion of existing correlations, similar to the work done by \citet{Gonzalez-Martin2018}, but the complexity of AGN X-ray emission and the lack of a large sample with high-quality data make it difficult to do. For black hole mass estimates, it is perhaps more appropriate to use observations taken when the source is at its high-flux state, for which the X-ray spectrum is generally {\it simple} (\citealt{Gallo2006}) and the X-ray variability mainly reflect the properties of the corona.
	
	\begin{table}[t!] 
		\begin{center}
			\caption{\edit1{{Black hole masses, luminosities and Eddington ratios of our sample.}} \label{tab:mass_of_sources}}
			\begin{tabular}{lccr}
				\hline
				Source Name        & $\log M_{{\rm BH}}$      & $\log L_{{\rm bol}}$ & $L_{\rm bol}/L_{\rm Edd}$\\
				\hline
				Mrk 335            & $7.23\pm 0.04^{\rm a}$   & $44.69^{\rm f}$      & 0.23 \\
				Mrk 766            & $6.2\pm 0.3^{\rm a}$     & $44.08^{\rm g}$      & 0.60 \\
				1H 0707-495        & $6.3\pm 0.5^{\rm a}$     & $44.43^{\rm h}$      & 1.1  \\
				IRAS 13224-3805    & $6.28\pm 0.05^{\rm b}$   & $45.55^{\rm h}$      & 15 \\
				NGC 1365           & $6.6\pm 0.3^{\rm c}$     & $43.8 ^{\rm i}$      & 0.13 \\
				NGC 3516           & $7.40\pm 0.05^{\rm a}$   & $44.29^{\rm f}$      & 0.062 \\
				NGC 4051           & $6.1\pm 0.1^{\rm a}$     & $43.56^{\rm f}$      & 0.23 \\
				NGC 4395           & $5.4\pm 0.2^{\rm d}$     & $41.0^{\rm d}$       & 0.0032 \\
				PG 1211+143        & $7.61\pm 0.15^{\rm e}$   & $45.81^{\rm f}$      & 1.3 \\
				\hline
			\end{tabular}
		\end{center}
		\vspace{-0.3cm} 
		\tablecomments{The BH masses of all the sources in out sample, bolometric luminosities \edit1{{and calculated Eddington ratios}} of the sources which have at least one detection of high-frequency break are listed. The values \edit1{{and errors}} of BH mass estimates in terms of ${\rm M_{\bigodot}}$ come from (a)\citet{Gonzalez-Martin2018} (which refers to \citet{Bian2003}, \citet{Bentz2009}, and \citet{Zu2011}), (b)\citet{Alston2020a},  (c)\citet{Combes2019}, (d)\edit1{{\citet{Brum2019}}}, and (e)\edit1{{\citet{Kaspi2000}}}.The values of bolometric luminosities in terms of ${\rm erg\ s^{-1}}$ are from (f)\citet{Woo2002},  \edit1{{(d)\citet{Brum2019}}}, (h)\citet{Zhou2005a}, and (i)\citet{Vasudevan2010}.}
		\vspace{0.2cm} 
	\end{table}

	\begin{table}[t!] 
		\begin{center}
			\caption{\edit1{{List of $\sigma_{\mathrm{rms}}^2$ computed in the 2–10 keV band with 20 ks intervals.}}\label{tab:rms20}}
            \begin{tabular}{lcr}
            \hline
            Source          & Obs     & $\sigma_{\mathrm{rms,\ 20ks}}^2$ \\ \hline
            Mrk 335         & 2006    & $0.011\pm 0.001$                 \\
                            & 2009(1) & $0.017\pm 0.006$                 \\
                            & 2009(2) & $0.020\pm 0.005$                 \\
            Mrk 766         & 2001    & $0.023\pm 0.002$                 \\
                            & 2005(1) & $0.032\pm 0.004$                 \\
                            & 2005(2) & $0.030\pm 0.002$                 \\
            1H 0707-495     & 2002    & $0.15\pm 0.04$                   \\
                            & 2007    & $0.23\pm 0.06$                   \\
                            & 2010    & $0.23\pm 0.07$                   \\
            IRAS 13224-3809 & 2002    & $0.19\pm 0.09$                   \\
                            & 2011    & $0.16\pm 0.10$                   \\
                            & 2016    & $0.25\pm 0.09$                   \\
            NGC 1365        & 2007    & $0.007\pm 0.007$                 \\
                            & 2012    & $0.035\pm 0.004$                 \\
                            & 2013    & $0.045\pm 0.003$                 \\
            NGC 3516        & 2001    & $(1.5\pm 1.0) \times 10^{-3}$    \\
                            & 2006    & $(7.0\pm 0.8) \times 10^{-3}$    \\
            NGC 4051        & 2001    & $0.105\pm 0.003$                   \\
                            & 2018    & $0.041\pm 0.003$                 \\
            NGC 4395        & 2003    & $0.16\pm 0.02$                   \\
                            & 2014    & $0.13\pm 0.01$                   \\
                            & 2019    & $0.12\pm 0.01$                   \\
            PG 1211+143     & 2001    & $(2.2\pm 3.9) \times 10^{-3}$    \\
                            & 2007    & $0.024\pm 0.006$                 \\ \hline
            \end{tabular}
		\end{center}
		\vspace{-0.3cm} 
		\vspace{0.2cm} 
	\end{table}

\section{Conclusions} \label{sec:6}
	In this work, we aim to explore the potential link between the X-ray PSD and spectral states for AGN. By making use of high-quality \textsl{XMM-Newton} data, we studied the spectral-timing properties of nine AGN exhibiting drastic spectral variation. We applied the Bayesian MCMC method for the PSD modelling, and generalized it for the comparison between different periodograms to identify the intrinsic PSD variation. Our results are summarized below.
	\begin{itemize}
	\itemsep0em
	\item We found that, out of the nine objects, four showed a high-frequency break in their PSDs in at least two of the observations. Among them, in three objects, namely, 1H 0707-495, Mrk 335 and Mrk 766, the break was found to exist only in a state with a high flux and a steep spectrum, but not in the low-flux state. Their rms variations were also found to be smaller in the high-flux state than low-flux. Interestingly, these sources are all NLS1s with high mass accretion rates near/over the Eddington limit.
	\item We found that NGC 4051 showed a PSD break in both high and low-flux states, and the break frequency in the low-flux state was higher.
    \item Based on the observations of the four AGN in which more than one PSD break is detected, we \edit1{{suggested}} a tendency that a large difference in the spectral flux and shape is associated with a large difference in the PSD shape. This result suggests a possible link between PSDs and spectral states in AGN.
	\item We investigated the effect of the PSD variation on the black hole mass estimate using X-ray variability for AGN. We found that the PSD variation could introduce as much as 1.0 dex variation to the X-ray rms, corresponding to $\sim$ 0.8 dex for the black hole mass estimate, although it is not the only reason to cause the dispersion observed in the $\sigma^2_{\rm rms}$-$M_{\rm BH}$ relation of \citet{Ponti2012}. Moreover, the change of the high-frequency break of the PSD results in a large scatter in the $f_{\rm B}$-$M_{\rm BH}$ relation, since the frequency $f_{\rm B}$ is likely depending on the spectral state. \edit1{{The correlation between timing properties and spectral state can provide a significant fraction of the uncertainty of these scaling relations to estimate the black hole mass for AGN.}}
	\end{itemize}
	
	It should be noted that our result is tentative given the limited sample size. \edit1{{Besides, the accretion rates of most of our sources are higher than a few per cent, and the frequency range investigated in this work has only two decades. Hence it is not possible to conduct a systematic comparison between our AGN sample and all the accretion states of XRBs.}} More observational data for a larger sample, especially taken at truly different accretion states, are essential to confirm this result. Besides, this work only covers the high-frequency break of AGN's PSD above $10^{-5}$ Hz. The study of the low-frequency break, like the work done for Ark 564 (\citealt{McHardy2007b}), requires long-term monitoring data covering the timescales from days to years. Future X-ray instruments of high-sensitivity and large field-of-view, such as the Einstein Probe mission (\citealt{Yuan2016}; \citealt{Yuan2018}), can provide monitoring data on timescales from days to years for hundreds of neighboring AGN. These future observations would, combining with deep observations of \textsl{XMM-Newton} and \textsl{eROSITA} (\citealt{Merloni2012}; \citealt{Predehl2016}), provide new datasets with which to improve the understanding of the link between the PSD and spectral state for AGN.

	\acknowledgments
	CJ acknowledges the National Natural Science Foundation of China through grant 11873054, as well as the support by the Strategic Pioneer Program on Space Science, Chinese Academy of Sciences through grant XDA15052100. This work has made use of observations conducted by \textit{XMM–Newton}, an ESA science mission with instruments and contributions directly funded by ESAMember States and the USA (NASA). This research has made use of the NASA/IPAC Extragalactic Database (NED) which is operated by the Jet Propulsion Laboratory, California Institute of Technology, under contract with the National Aeronautics and Space Administration.
	
    \facilities{{\it XMM-Newton}}
    
    \software{
    emcee \citep{Foreman-Mackey2013},
    HEAsoft \citep{NASA2014},
    matplotlib \citep{Hunter2007},
    numpy \citep{vanderWalt2011},
    scipy \citep{Virtanen2020},
    SAOImage DS9 \citep{Joye2003},
    SAS \citep{Gabriel2004},
    XSPEC \citep{Arnaud1996}
    }

	\newpage
	\bibliography{MSyang}{}
	\bibliographystyle{aasjournal}

	\appendix
	\setcounter{table}{0}   
	\setcounter{figure}{0}
	\renewcommand{\thetable}{A\arabic{table}}
	\renewcommand{\thefigure}{A\arabic{figure}}

	\section{Figures and Tables} \label{appendix:figures&tables}

	\begin{figure*}[ht!] 
		\begin{center} 
			\gridline{
				\fig{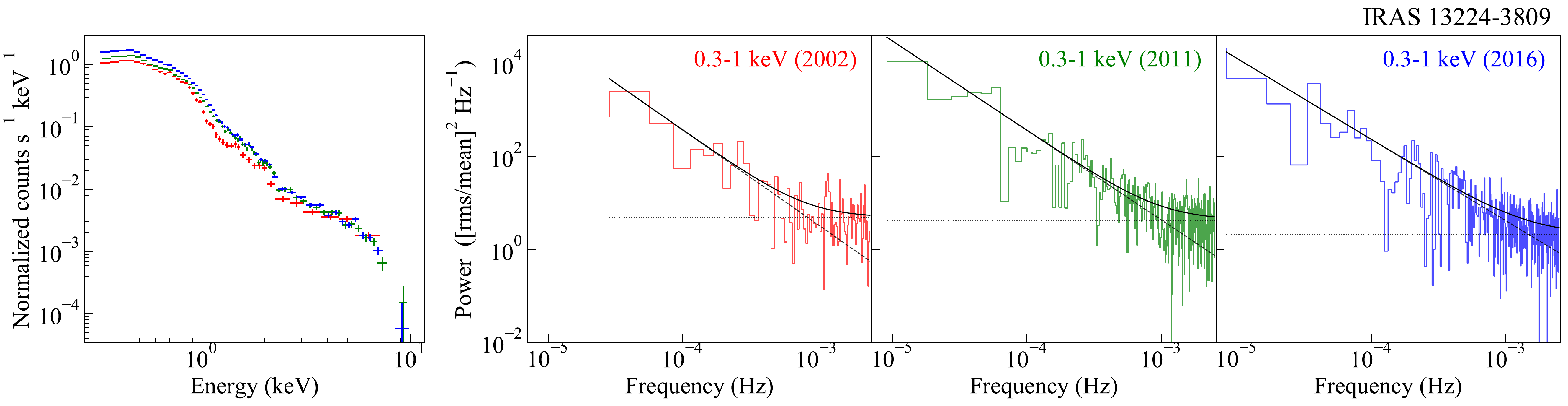}{1.0\textwidth}{}
			}
			\vspace{-0.5cm}
			\gridline{
				\fig{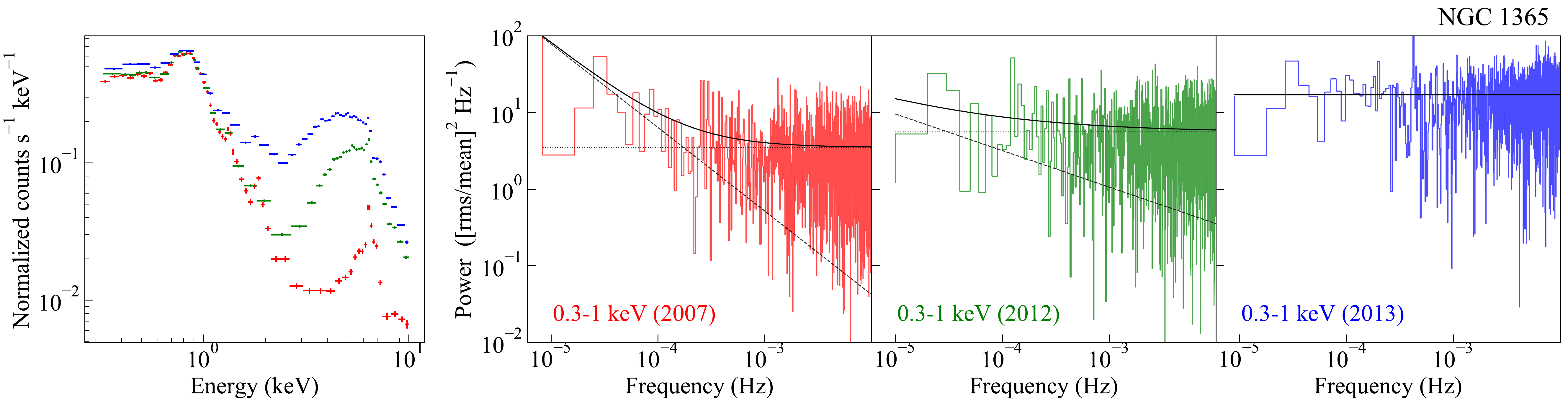}{1.0\textwidth}{}
			}
			\vspace{-0.5cm}
			\gridline{
				\fig{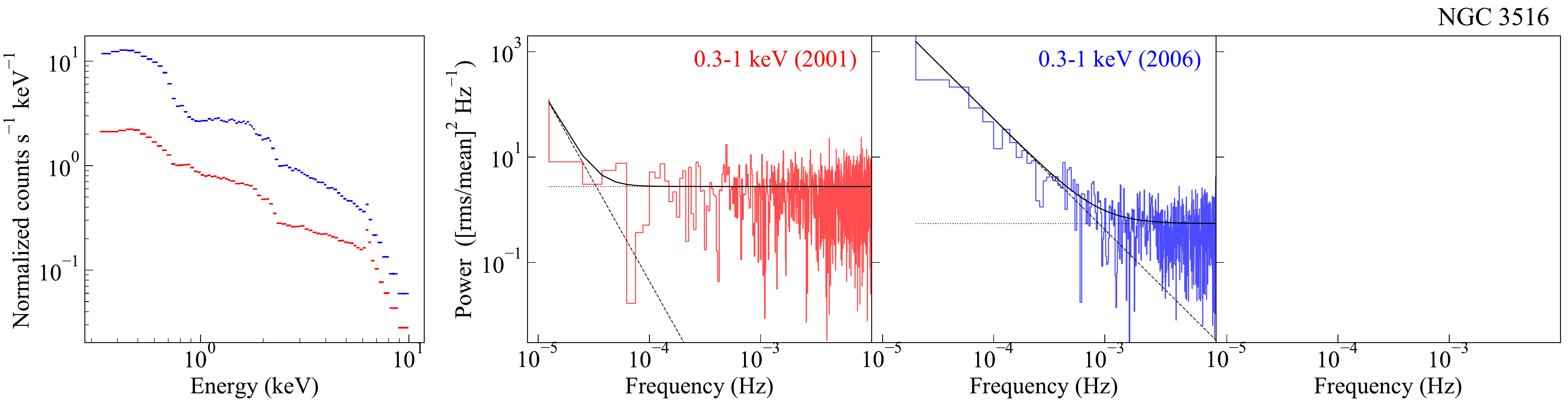}{1.0\textwidth}{}
			}
			\vspace{-0.5cm}
			\caption{The energy spectra and 0.3-1 keV PSDs of the AGN which have no more than two detection of high-frequency break. Black lines are the best-fit Model-B (i.e. break power law) results for observations which report significant breaks, and are the best-fit Model-A (i.e. single power law) results for observations which do not report significant breaks. The dashed lines and the dotted lines represent red noise components and Poisson terms, respectively. Particularly, the LRT1 \textit{p}-value of Model-A simulation for the 2001 observation of PG 1211+143 is shown downside since it is lower than 0.01. Spectra and PSDs are shown in different colors attending to the observation date. The PSDs are arranged in order of fluxes. \label{fig:spec&PSD_unused_s}}
		\end{center}
	\end{figure*}
	\addtocounter{figure}{-1} 
	\begin{figure*}
		\begin{center} 
			\gridline{
				\fig{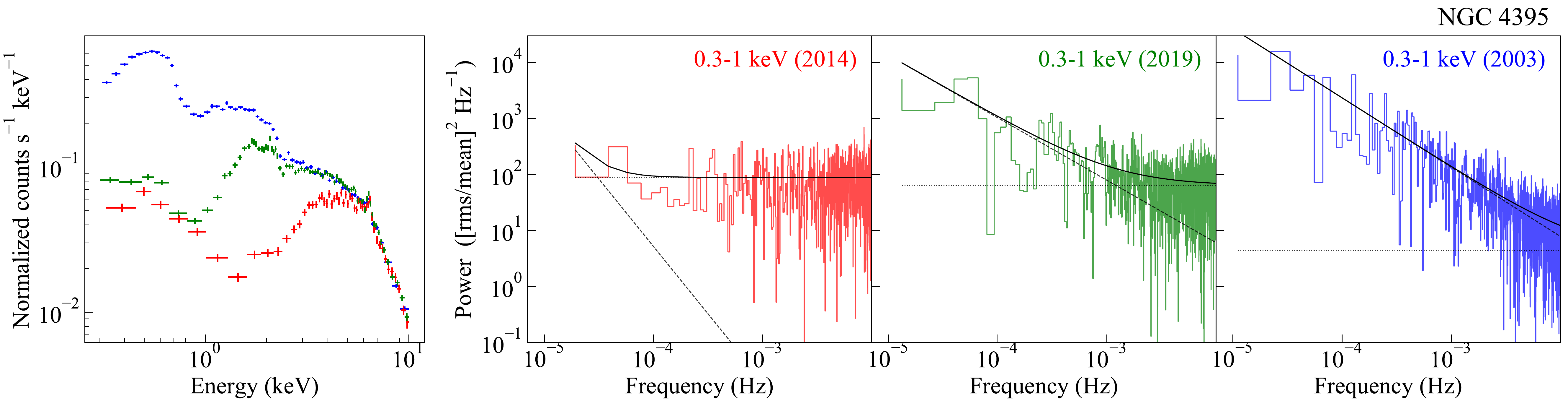}{1.0\textwidth}{}
			}
			\vspace{-0.5cm}
			\gridline{
				\fig{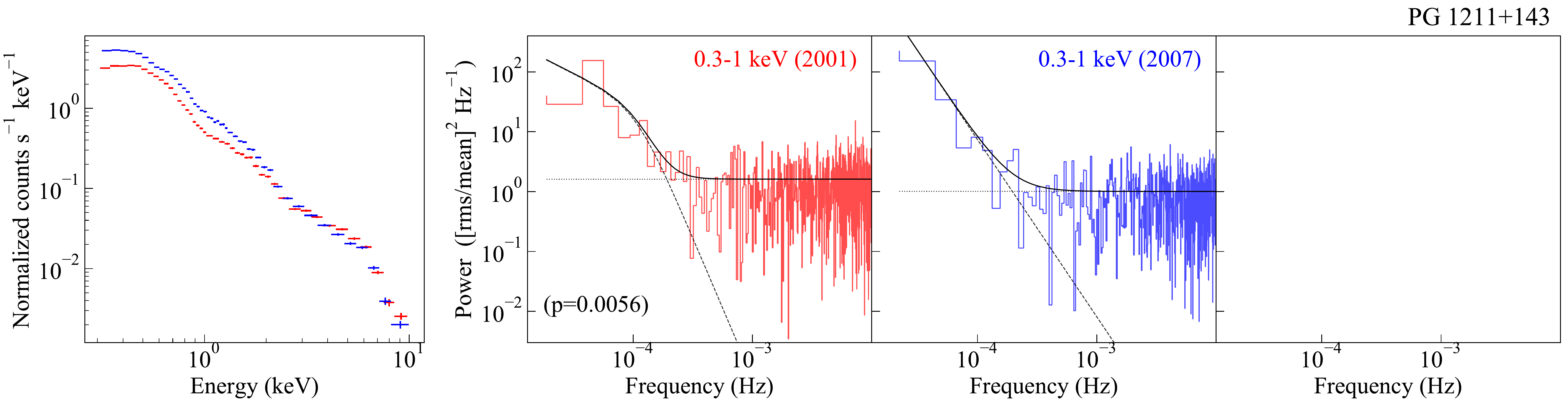}{1.0\textwidth}{}
			}
			\vspace{-0.5cm}
			\caption{continued}
		\end{center} 
	\end{figure*}
	\clearpage
	
	\begin{figure*}[ht!] 
		\begin{center} 
			\gridline{
				\fig{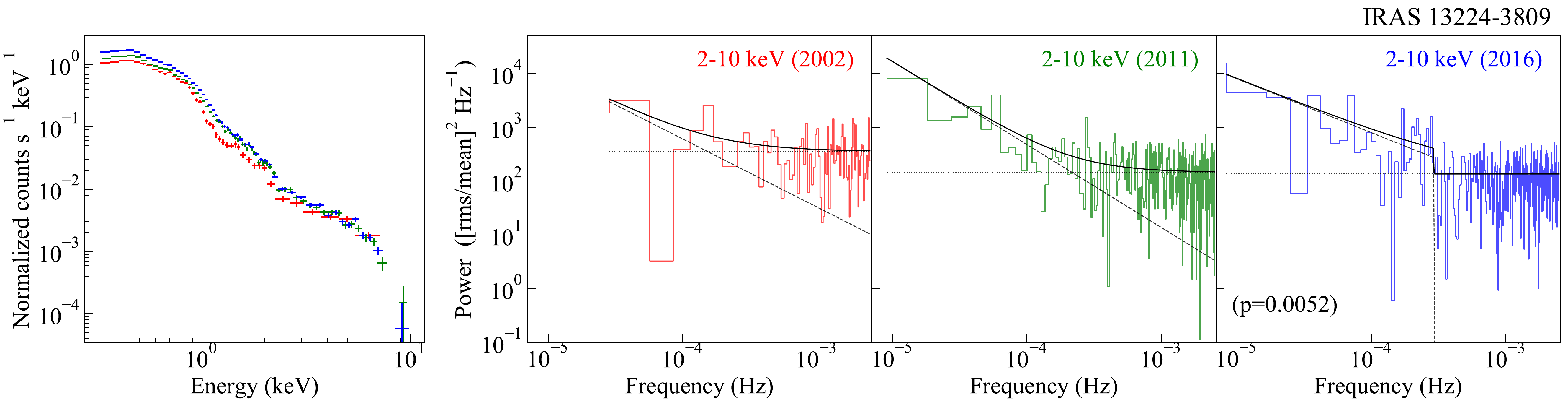}{1.0\textwidth}{}
			}
			\vspace{-0.5cm}
			\gridline{
				\fig{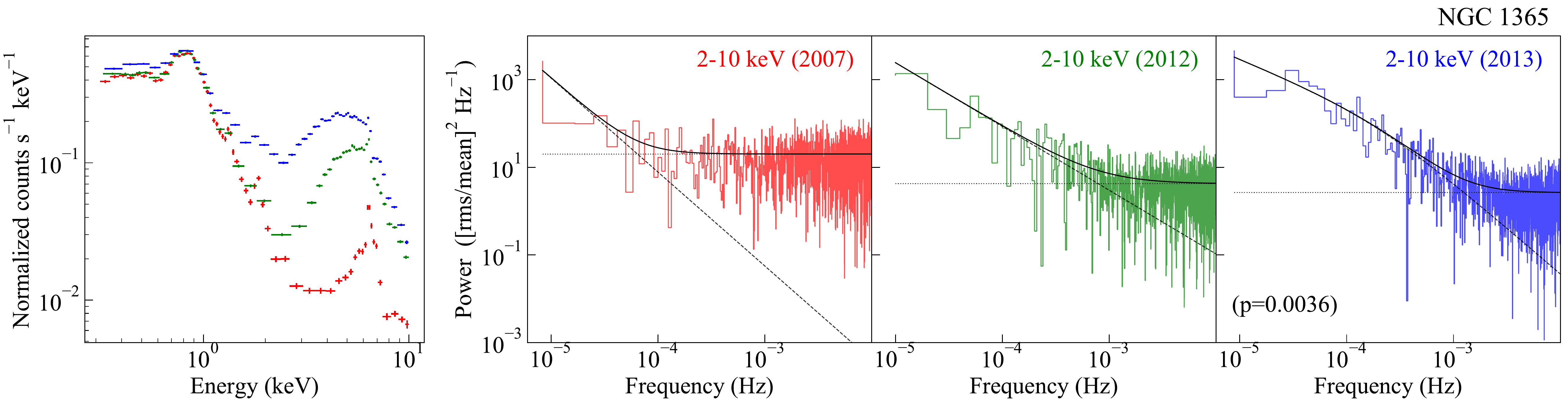}{1.0\textwidth}{}
			}
			\vspace{-0.5cm}
			\gridline{
				\fig{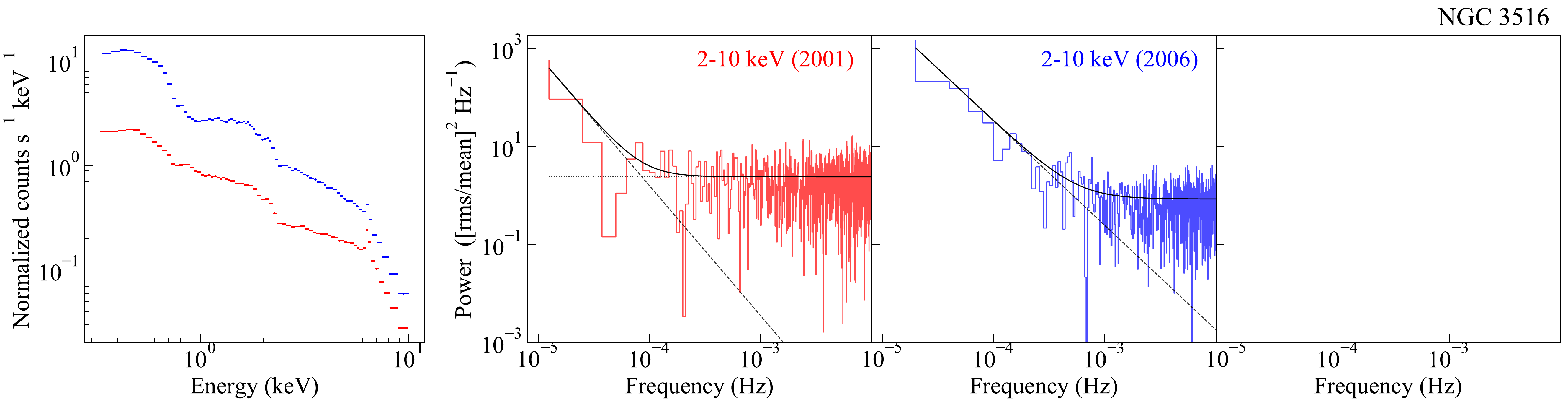}{1.0\textwidth}{}
			}
			\vspace{-0.5cm}
			\gridline{
				\fig{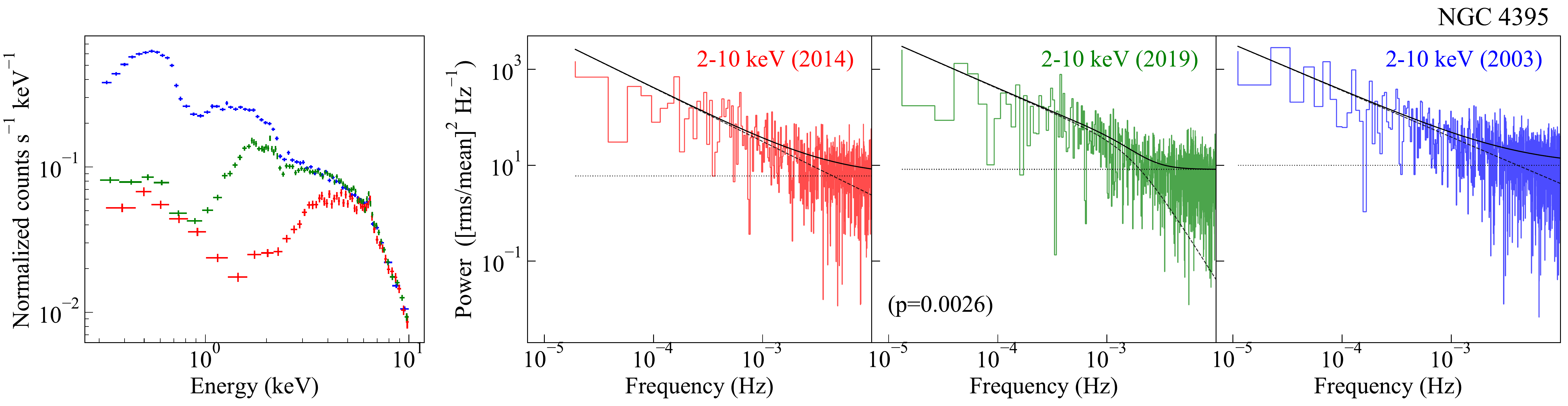}{1.0\textwidth}{}
			}
			\vspace{-0.5cm}
			\caption{The energy spectra and 0.3-1 keV PSDs of the AGN which have only one or no detection of high-frequency break. Black lines are the best-fit Model-B (i.e. break power law) results for observations which report significant breaks, and are the best-fit Model-A (i.e. single power law) results for observations which do not report significant breaks. The dashed lines and the dotted lines represent red noise components and Poisson terms, respectively. Particularly, the LRT1 \textit{p}-values of Model-A simulation for the 2016 observation of IRAS 13224-3809, the 2013 observation of NGC 1365, and the 2019 observation of NGC 4395 are shown downside since they are lower than 0.01. Spectra and PSDs are shown in different colors attending to the observation date. The PSDs are arranged in order of fluxes. \label{fig:spec&PSD_unused_h}}
		\end{center}
	\end{figure*}
	\addtocounter{figure}{-1} 
	\begin{figure*}
		\begin{center} 
			\gridline{
				\fig{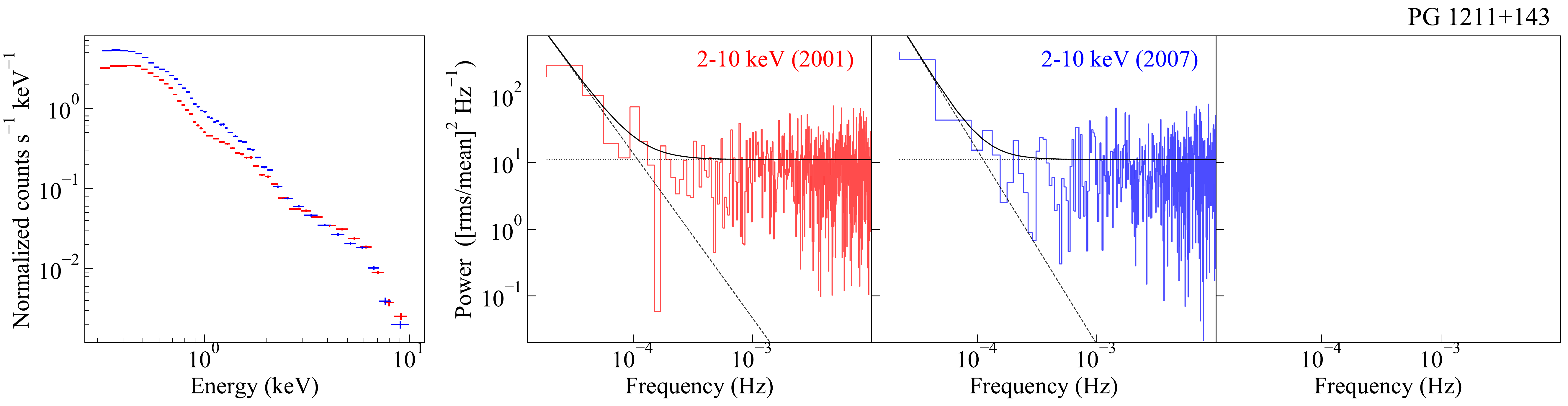}{1.0\textwidth}{}
			}
			\vspace{-0.5cm}
			\caption{continued}
		\end{center} 
	\end{figure*}
	\clearpage

	\begin{figure*}[ht!] 
		\begin{center} 
			\gridline{\fig{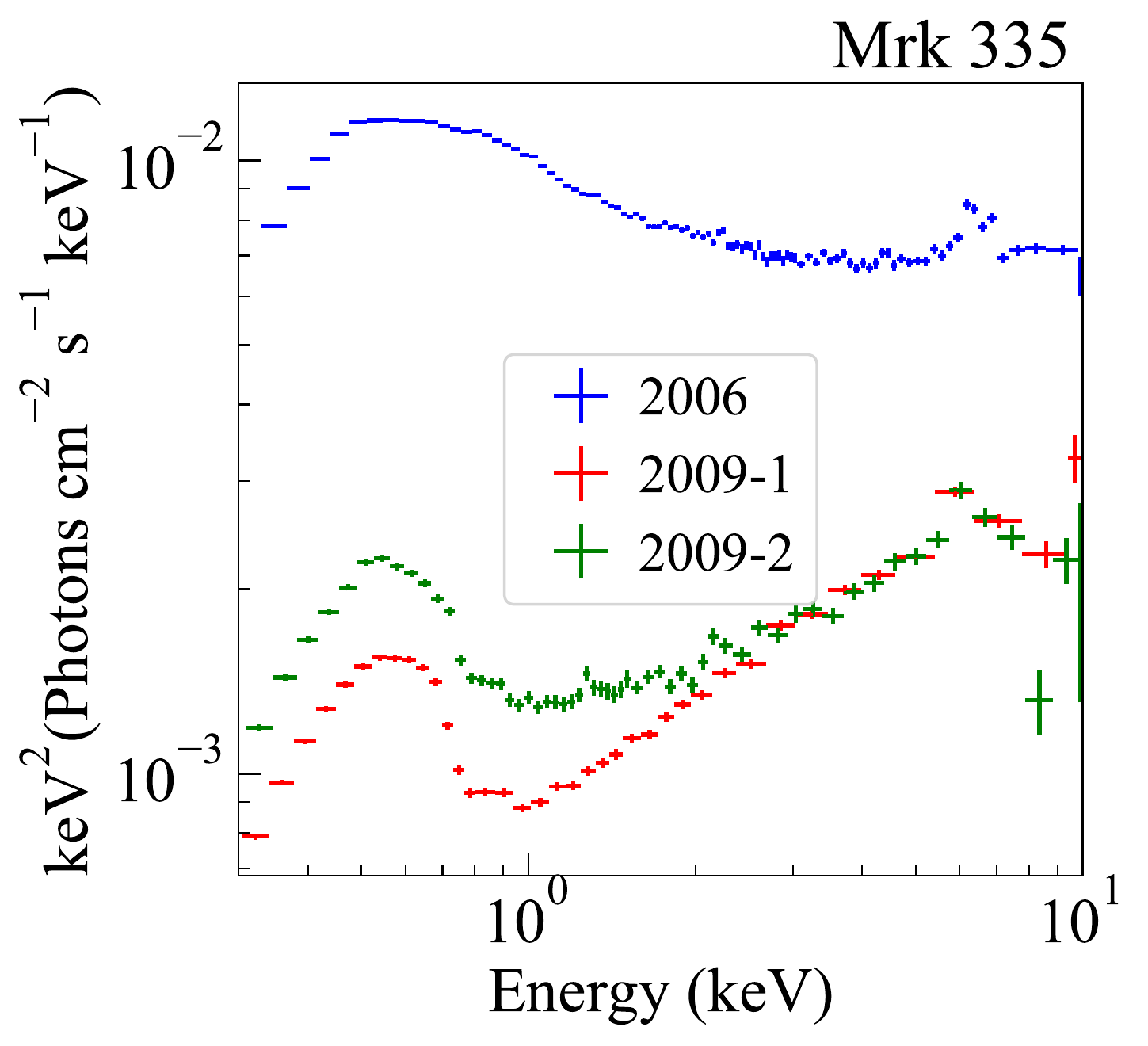}{0.3\textwidth}{}
			    \fig{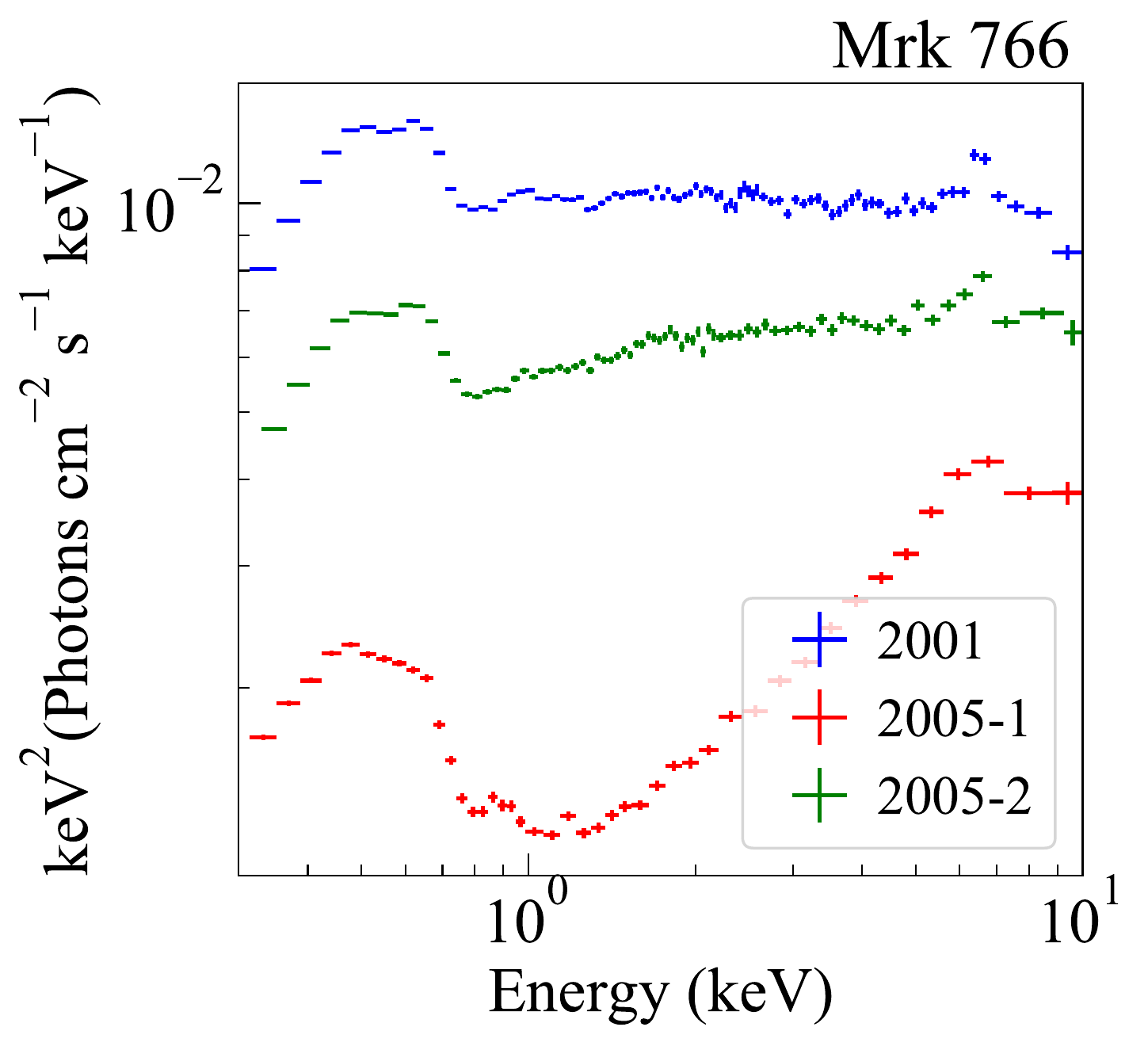}{0.3\textwidth}{}
			    \fig{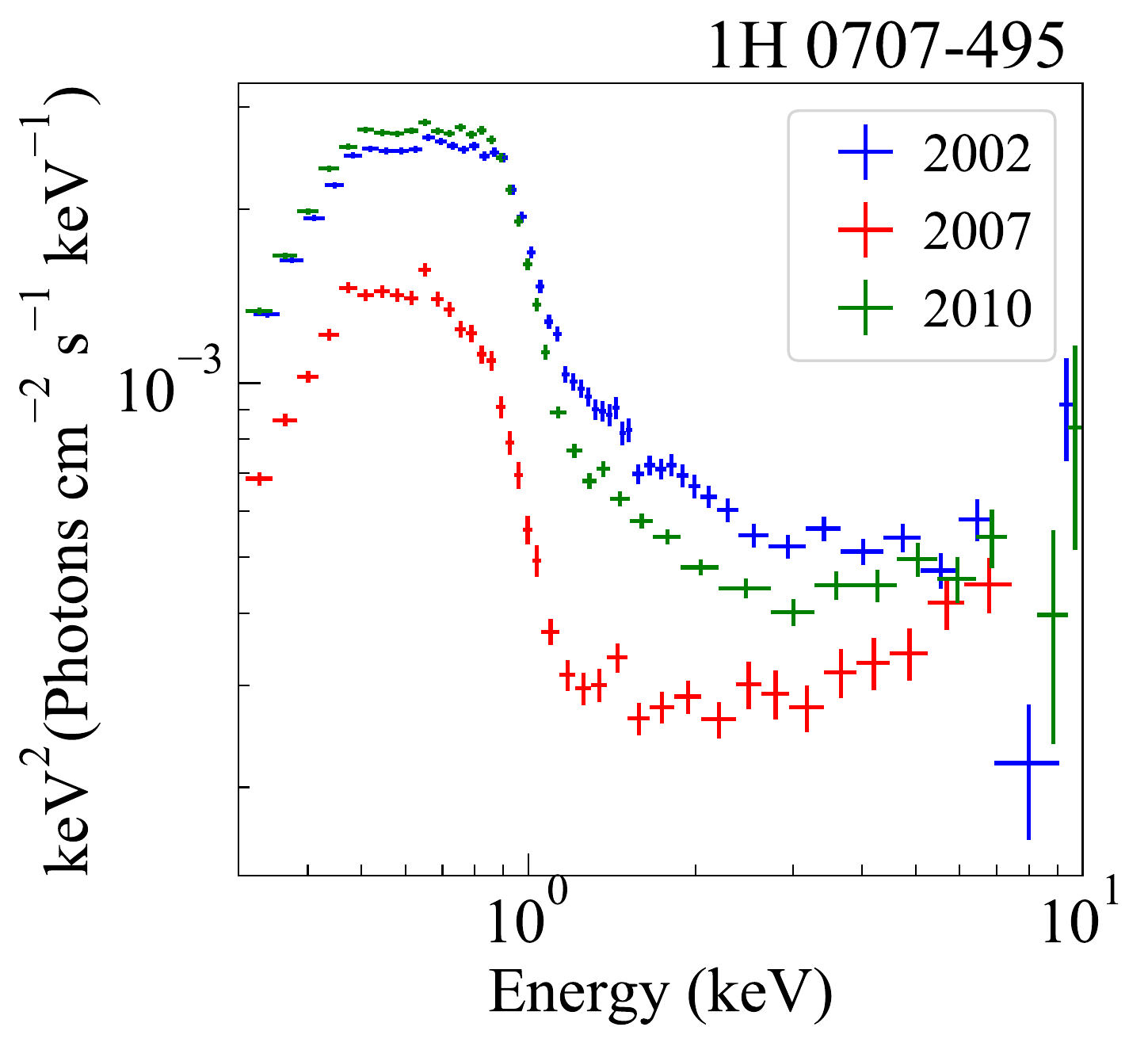}{0.3\textwidth}{}
			}
			\vspace{-0.6cm}
			\gridline{\fig{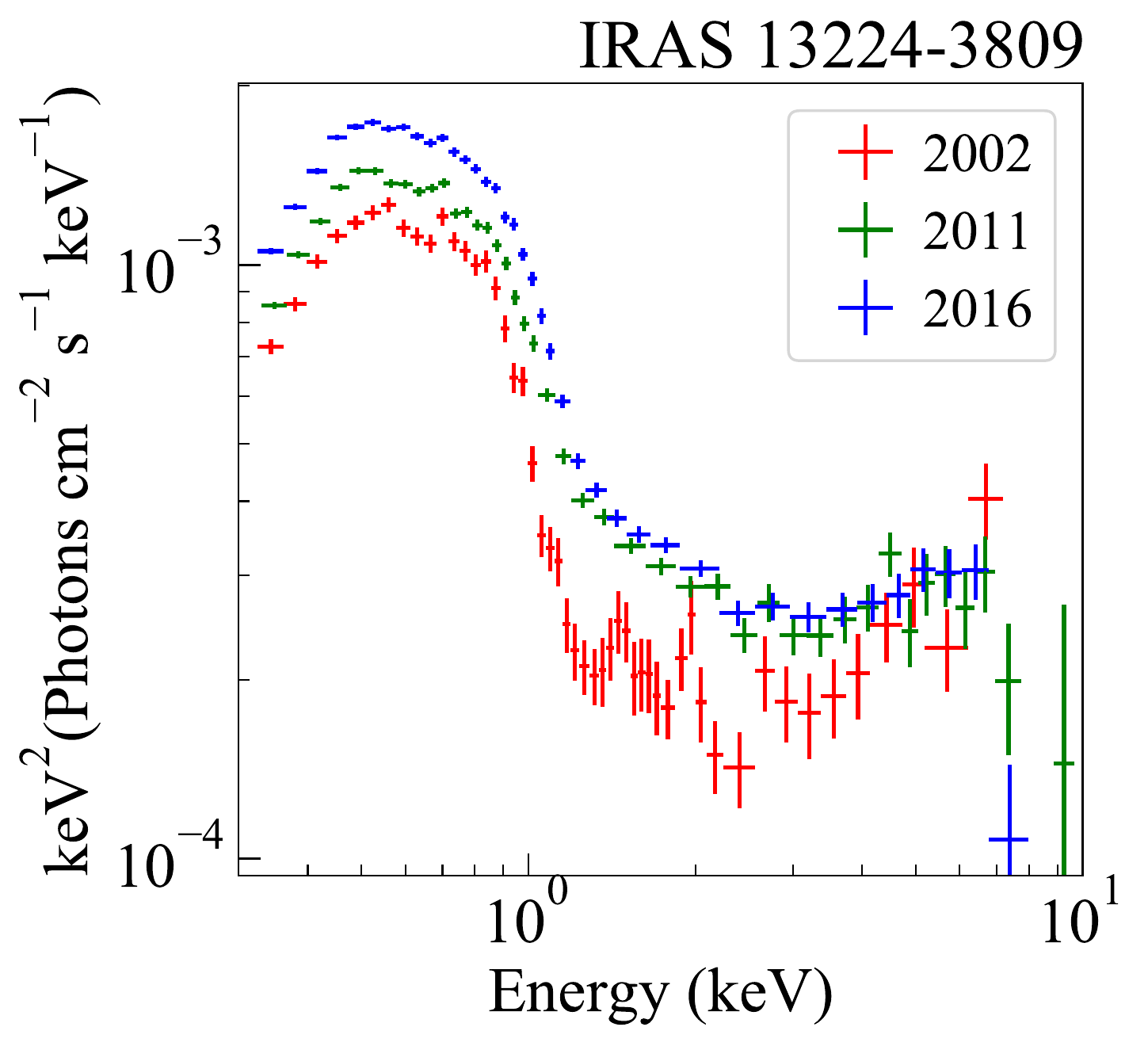}{0.3\textwidth}{}
				\fig{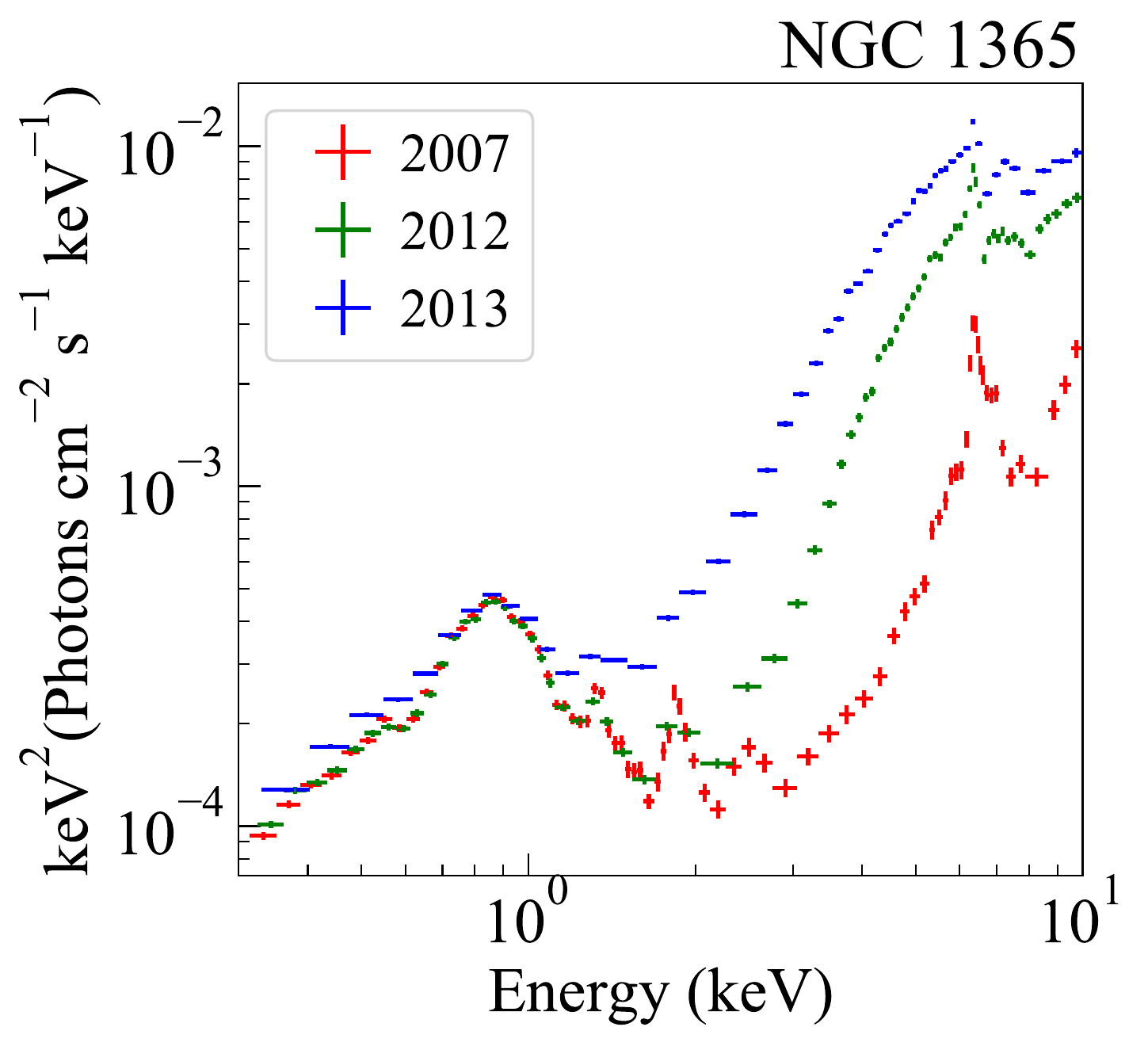}{0.3\textwidth}{}
				\fig{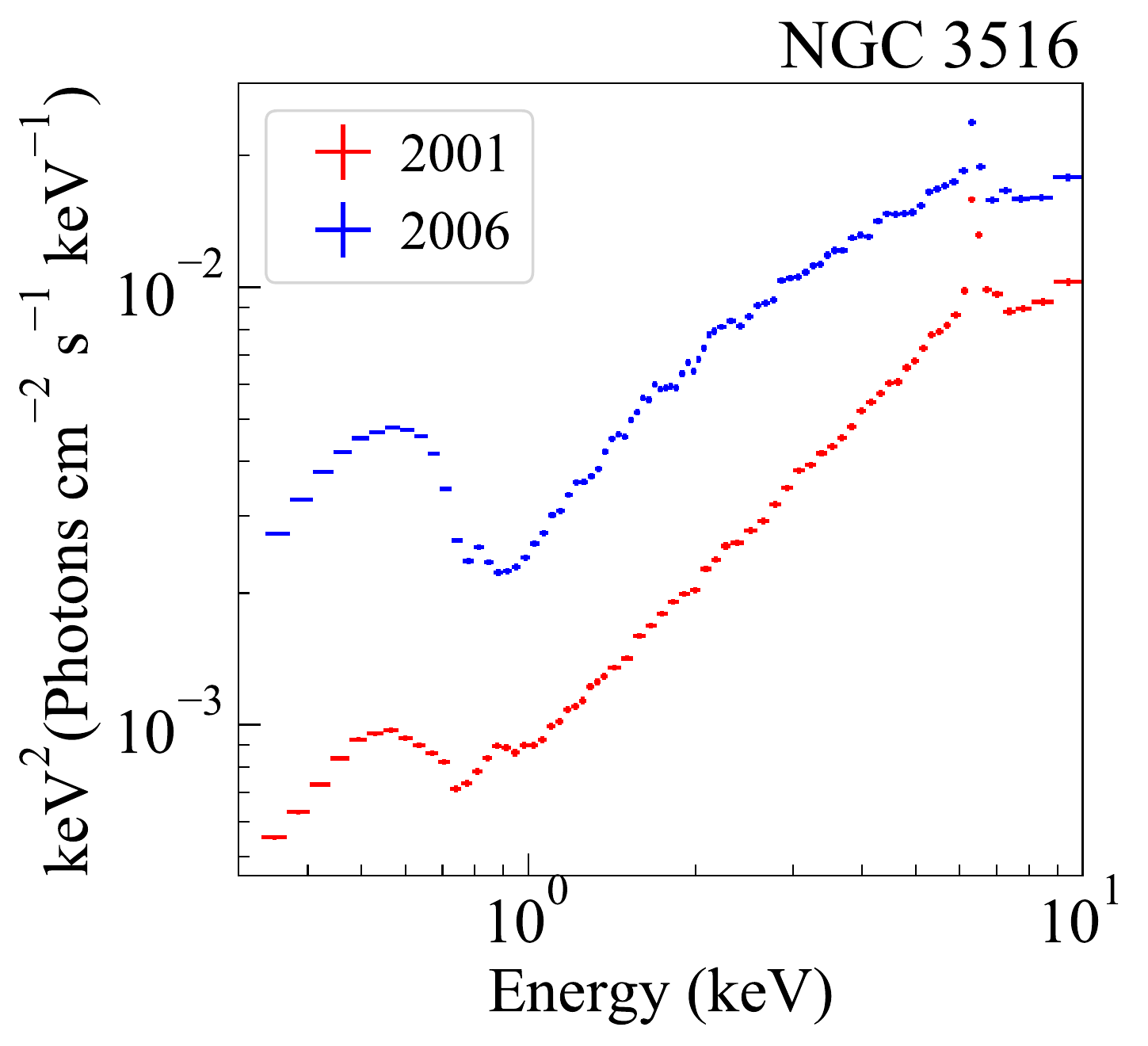}{0.3\textwidth}{}
			}
			\vspace{-0.6cm}
			\gridline{\fig{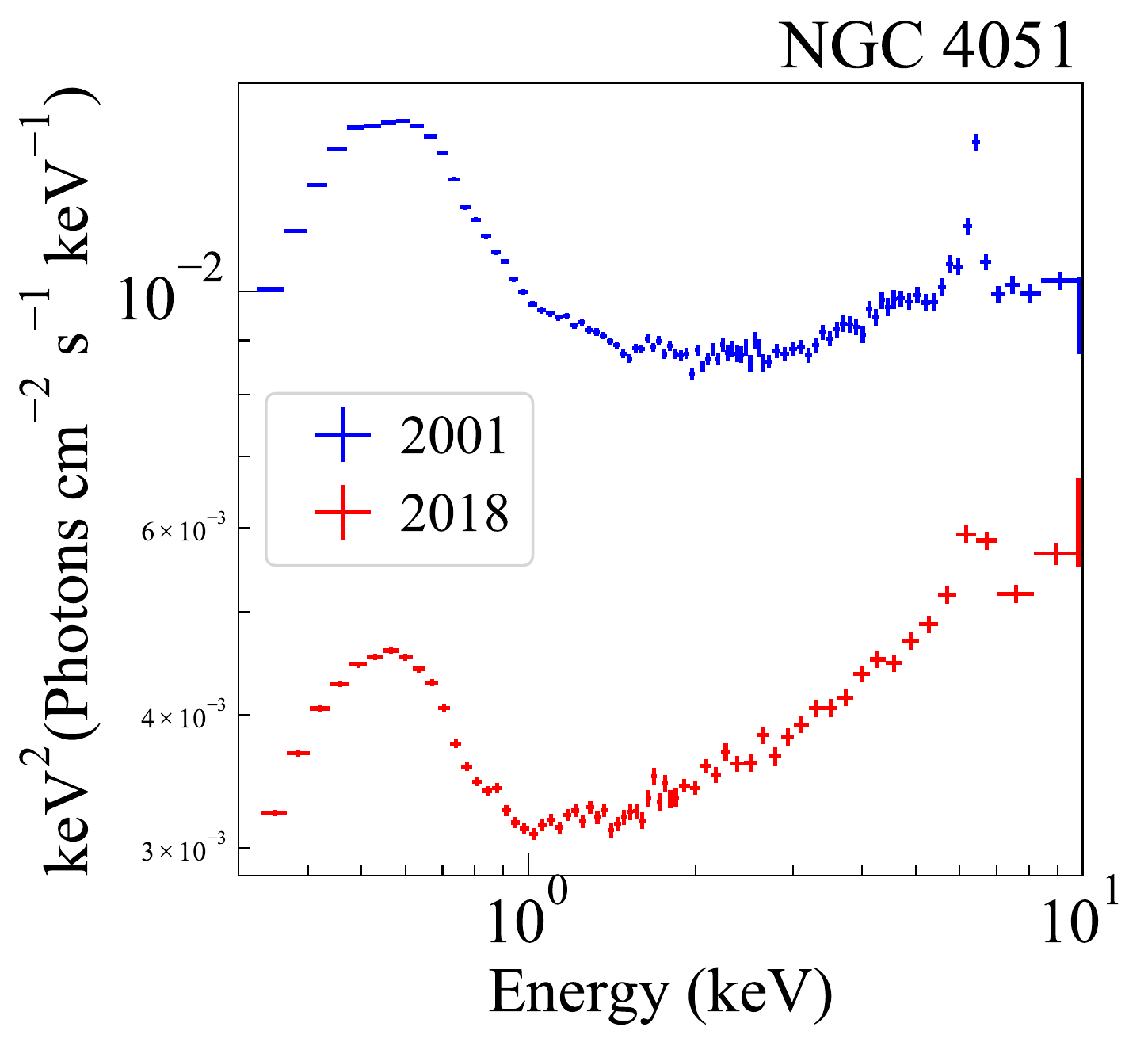}{0.3\textwidth}{}
				\fig{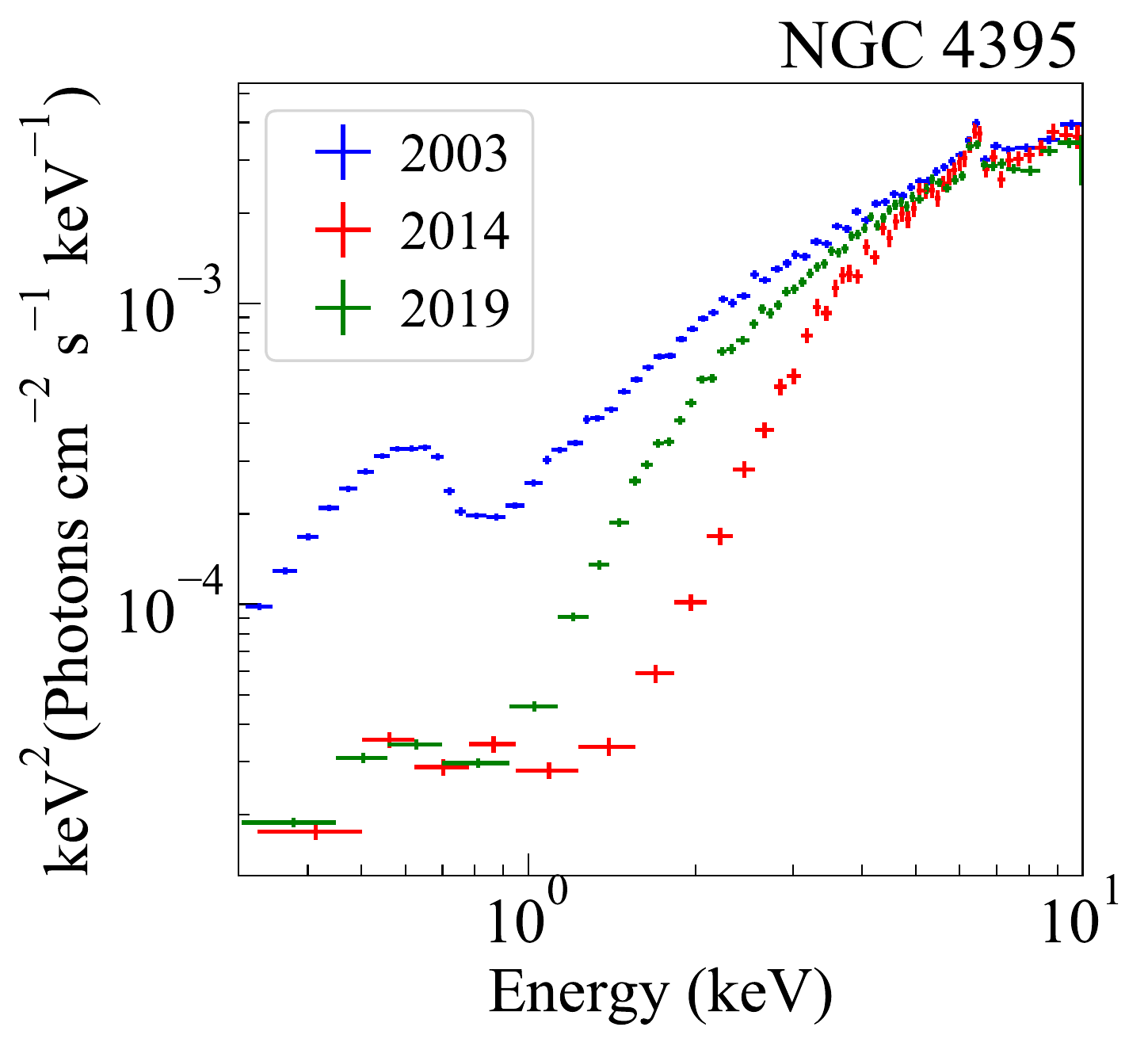}{0.3\textwidth}{}
				\fig{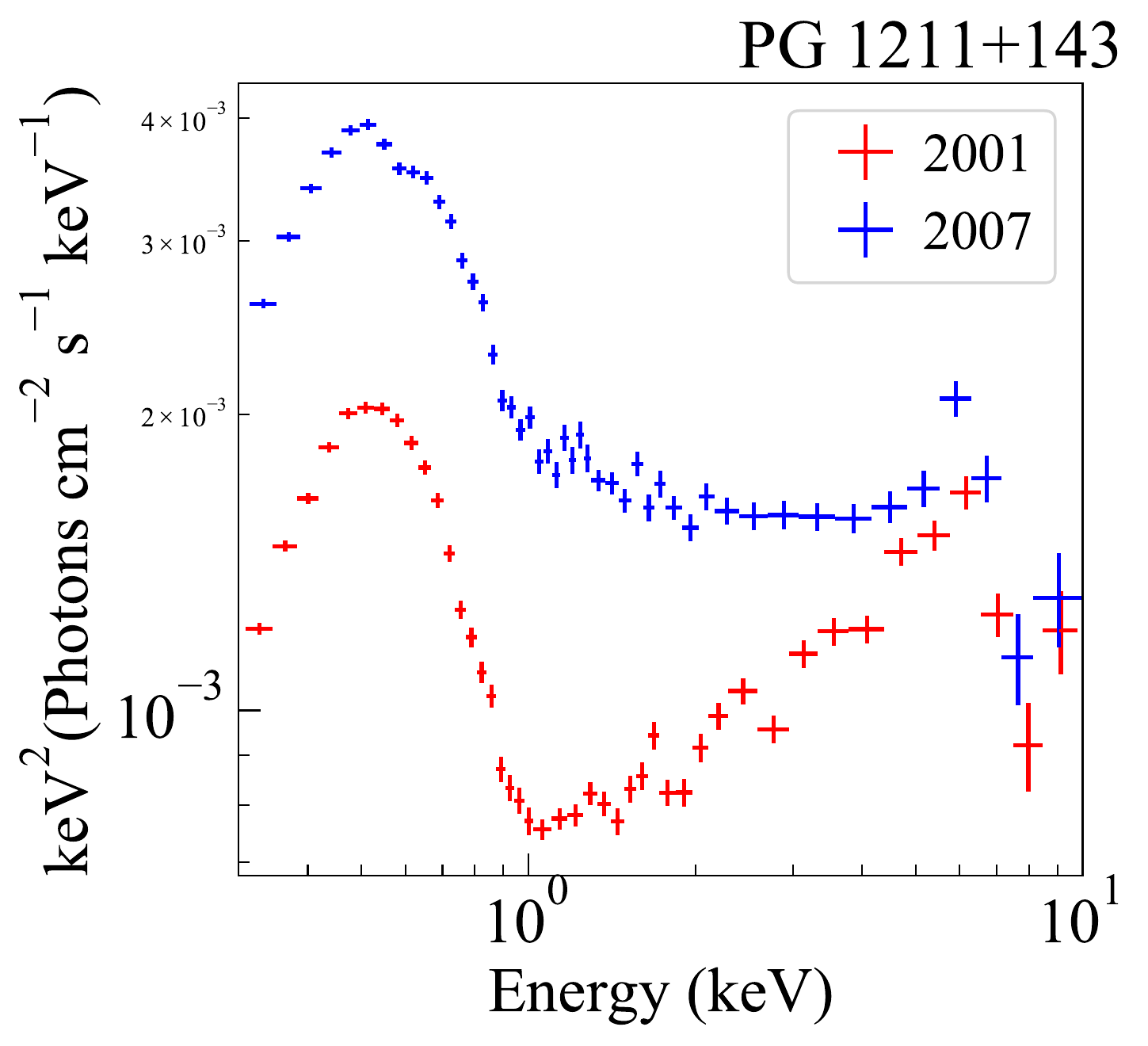}{0.3\textwidth}{}
			}
			\vspace{-0.6cm}
			\caption{\edit3{{Time-averaged `fluxed' spectra unfolded through a power law model with index 0 and normalization 1, which are largely free from effects of energy-dependent effective area and give representations of the underlying source spectra.}} \label{fig:unfoldspec}}
		\end{center}
	\end{figure*}
	\clearpage

	\begin{table*}
		\renewcommand\arraystretch{1.2}
		\begin{center}
			\caption{\edit1{{Parameters of the PSD fits to Model A}} \label{tab:parameters_modelA}}
			\begin{tabular}{lcrrrrrr}
                \hline
                Source          & Obs     & \multicolumn{3}{c}{0.3-1 keV}                                               & \multicolumn{3}{c}{2-10 keV}                                                \\
                &         & $\log(A)$                & $\alpha$               & $\log(C)$               & $\log(A)$                & $\alpha$               & $\log(C)$               \\ \hline
                Mrk 335         & 2006    & $-8.62_{-0.51}^{+0.49}$  & $2.52_{-0.14}^{+0.15}$ & $-0.78_{-0.01}^{+0.01}$ & $-6.66_{-0.64}^{+0.59}$  & $2.09_{-0.17}^{+0.17}$ & $0.10_{-0.01}^{+0.01}$  \\
                & 2009(1) & $-7.90_{-0.71}^{+0.65}$  & $2.41_{-0.17}^{+0.19}$ & $0.21_{-0.01}^{+0.01}$  & $-4.82_{-0.89}^{+0.76}$  & $1.65_{-0.19}^{+0.23}$ & $1.04_{-0.02}^{+0.02}$  \\
                & 2009(2) & $-8.82_{-0.99}^{+0.87}$  & $2.63_{-0.22}^{+0.25}$ & $0.29_{-0.02}^{+0.02}$  & $-5.38_{-1.45}^{+1.09}$  & $1.76_{-0.27}^{+0.35}$ & $1.01_{-0.02}^{+0.02}$  \\
                Mrk 766         & 2001    & $-6.89_{-0.37}^{+0.36}$  & $2.25_{-0.11}^{+0.12}$ & $-0.75_{-0.02}^{+0.02}$ & $-4.40_{-0.39}^{+0.37}$  & $1.58_{-0.11}^{+0.12}$ & $-0.01_{-0.03}^{+0.03}$ \\
                & 2005(1) & $-5.97_{-0.58}^{+0.51}$  & $1.92_{-0.15}^{+0.16}$ & $0.02_{-0.02}^{+0.02}$  & $-4.22_{-0.49}^{+0.45}$  & $1.56_{-0.13}^{+0.14}$ & $0.55_{-0.02}^{+0.02}$  \\
                & 2005(2) & $-6.77_{-0.48}^{+0.46}$  & $2.20_{-0.14}^{+0.15}$ & $-0.46_{-0.02}^{+0.02}$ & $-4.92_{-0.53}^{+0.48}$  & $1.72_{-0.14}^{+0.16}$ & $-0.14_{-0.02}^{+0.02}$ \\
                1H 0707-495     & 2002    & $-4.48_{-0.74}^{+0.71}$  & $1.69_{-0.20}^{+0.21}$ & $0.02_{-0.53}^{+0.18}$  & $-2.70_{-1.12}^{+1.00}$  & $1.32_{-0.26}^{+0.29}$ & $1.87_{-0.07}^{+0.05}$  \\
                & 2007    & $-4.93_{-1.01}^{+0.75}$  & $1.96_{-0.22}^{+0.28}$ & $0.26_{-0.77}^{+0.25}$  & $-1.90_{-1.38}^{+1.29}$  & $1.21_{-0.33}^{+0.37}$ & $1.72_{-0.76}^{+0.15}$  \\
                & 2010    & $-6.41_{-0.64}^{+0.61}$  & $2.28_{-0.17}^{+0.18}$ & $0.37_{-0.08}^{+0.06}$  & $-3.41_{-0.93}^{+0.84}$  & $1.59_{-0.22}^{+0.25}$ & $2.00_{-0.05}^{+0.04}$  \\
                IRAS 13224-3809 & 2002    & $-5.48_{-1.35}^{+1.30}$  & $2.01_{-0.35}^{+0.37}$ & $0.68_{-0.19}^{+0.11}$  & $-5.70_{-15.18}^{+4.48}$ & $1.89_{-0.99}^{+2.19}$ & $2.59_{-0.09}^{+0.07}$  \\
                & 2011    & $-5.15_{-0.50}^{+0.47}$  & $1.93_{-0.13}^{+0.14}$ & $0.63_{-0.08}^{+0.07}$  & $-3.93_{-1.47}^{+1.16}$  & $1.64_{-0.27}^{+0.35}$ & $2.17_{-0.05}^{+0.04}$  \\
                & 2016    & $-4.57_{-0.46}^{+0.42}$  & $1.73_{-0.11}^{+0.13}$ & $0.31_{-0.13}^{+0.10}$  & $-2.23_{-0.79}^{+0.72}$  & $1.26_{-0.19}^{+0.20}$ & $2.03_{-0.07}^{+0.05}$  \\
                NGC 1365        & 2007    & $-3.84_{-1.03}^{+0.81}$  & $1.16_{-0.20}^{+0.25}$ & $0.55_{-0.02}^{+0.02}$  & $-8.96_{-3.29}^{+2.22}$  & $2.42_{-0.49}^{+0.70}$ & $1.30_{-0.01}^{+0.01}$  \\
                & 2012    & $-3.90_{-4.14}^{+2.79}$  & $0.66_{-0.43}^{+0.62}$ & $0.78_{-0.05}^{+0.02}$  & $-3.95_{-0.46}^{+0.43}$  & $1.47_{-0.12}^{+0.13}$ & $0.63_{-0.02}^{+0.02}$  \\
                & 2013    & $-9.17^*$                & $1.13^*$               & $1.23_{-0.01}^{+0.01}$  & $-4.45_{-0.32}^{+0.30}$  & $1.69_{-0.09}^{+0.09}$ & $0.41_{-0.02}^{+0.02}$  \\
                NGC 3516        & 2001    & $-16.38^*$               & $3.76^*$               & $0.44_{-0.02}^{+0.02}$  & $-15.05_{-8.80}^{+5.01}$ & $3.66_{-1.08}^{+1.90}$ & $0.38_{-0.02}^{+0.02}$  \\
                & 2006    & $-6.78_{-0.62}^{+0.56}$  & $2.13_{-0.16}^{+0.18}$ & $-0.26_{-0.02}^{+0.02}$ & $-7.17_{-0.85}^{+0.74}$  & $2.17_{-0.20}^{+0.23}$ & $-0.07_{-0.02}^{+0.02}$ \\
                NGC 4051        & 2001    & $-5.18_{-0.18}^{+0.17}$  & $1.99_{-0.06}^{+0.06}$ & $-0.99_{-0.09}^{+0.07}$ & $-3.40_{-0.23}^{+0.23}$  & $1.44_{-0.07}^{+0.07}$ & $-0.01_{-0.06}^{+0.05}$ \\
                & 2018    & $-4.82_{-0.36}^{+0.34}$  & $1.90_{-0.12}^{+0.12}$ & $-0.50_{-0.08}^{+0.07}$ & $-3.57_{-0.49}^{+0.46}$  & $1.45_{-0.14}^{+0.15}$ & $0.31_{-0.05}^{+0.04}$  \\
                NGC 4395        & 2003    & $-1.42_{-0.28}^{+0.22}$  & $1.19_{-0.07}^{+0.09}$ & $0.45_{-1.17}^{+0.28}$  & $-1.28_{-0.32}^{+0.31}$  & $0.96_{-0.09}^{+0.10}$ & $0.99_{-0.11}^{+0.07}$  \\
                & 2014    & $-8.91^*$                & $2.41^*$               & $1.95_{-0.02}^{+0.02}$  & $-1.82_{-0.43}^{+0.42}$  & $1.11_{-0.13}^{+0.14}$ & $0.77_{-0.14}^{+0.09}$  \\
                & 2019    & $-1.48_{-0.46}^{+0.43}$  & $1.12_{-0.13}^{+0.13}$ & $1.81_{-0.04}^{+0.03}$  & $-2.24_{-0.34}^{+0.33}$  & $1.23_{-0.10}^{+0.11}$ & $0.74_{-0.08}^{+0.06}$  \\
                PG 1211+143     & 2001    & $-9.82_{-2.23}^{+1.85}$  & $2.71_{-0.46}^{+0.56}$ & $0.21_{-0.02}^{+0.02}$  & $-10.23_{-4.07}^{+2.83}$ & $2.82_{-0.68}^{+0.96}$ & $1.05_{-0.02}^{+0.02}$  \\
                & 2007    & $-11.83_{-2.84}^{+2.17}$ & $3.17_{-0.54}^{+0.69}$ & $0.00_{-0.02}^{+0.02}$  & $-12.48_{-5.34}^{+3.69}$ & $3.40_{-0.88}^{+1.25}$ & $1.05_{-0.02}^{+0.02}$  \\ \hline
            \end{tabular}
		\end{center}
		\tablecomments{*The soft band observations of NGC 1365 in 2013, NGC 3516 in 2001, and NGC 4395 in 2014 are dominated by Poisson noise. The other two parameters are poorly constrained and only best-fit values are shown here.}
	\end{table*}

	\begin{table*}
		\renewcommand\arraystretch{1.2}
		\begin{center}
			\caption{\edit1{{Parameters of the PSD fits to Model B}} \label{tab:parameters_modelB}}
			\begin{tabular}{lcrrrrrrrr}
            \hline
            Source      & Obs     & \multicolumn{4}{c}{0.3-1 keV}                                                                        & \multicolumn{4}{c}{2-10 keV}                                                                         \\
            &         & $\log(A)$               & $\log(f_{\mathrm{B}})$  & $\alpha_{\mathrm{H}}$  & $\log(C)$               & $\log(A)$               & $\log(f_{\mathrm{B}})$  & $\alpha_{\mathrm{H}}$  & $\log(C)$               \\ \hline
            Mrk 335     & 2006    & $−2.27_{-0.15}^{+0.20}$ & $-3.88_{-0.17}^{+0.13}$ & $3.31_{-0.33}^{+0.40}$ & $-0.77_{-0.01}^{+0.01}$ & $−2.25_{-0.12}^{+0.14}$ & $-3.63_{-0.13}^{+0.10}$ & $3.83_{-0.66}^{+0.95}$ & $0.12_{-0.01}^{+0.01}$  \\
            & 2009(1) & -                       & -                       & -                      & -                       & -                       & -                       & -                      & -                       \\
            & 2009(2) & -                       & -                       & -                      & -                       & -                       & -                       & -                      & -                       \\
            Mrk 766     & 2001    & $−2.16_{-0.11}^{+0.13}$ & $-3.38_{-0.13}^{+0.10}$ & $3.42_{-0.39}^{+0.49}$ & $-0.71_{-0.02}^{+0.02}$ & $−2.16_{-0.10}^{+0.15}$ & $-3.14_{-0.24}^{+0.11}$ & $3.35_{-0.99}^{+1.67}$ & $0.05_{-0.03}^{+0.02}$  \\
            & 2005(1) & -                       & -                       & -                      & -                       & -                       & -                       & -                      & -                       \\
            & 2005(2) & $−2.25_{-0.10}^{+0.13}$ & $-3.31_{-0.13}^{+0.09}$ & $4.47_{-0.92}^{+1.43}$ & $-0.43_{-0.02}^{+0.02}$ & $−2.21_{-0.08}^{+0.09}$ & $-3.16_{-0.06}^{+0.04}$ & $8.53_{-3.31}^{+9.58}$ & $0.19_{-0.02}^{+0.02}$  \\
            1H 0707-495 & 2002    & $-1.65_{-0.14}^{+0.22}$ & $-3.54_{-0.30}^{+0.16}$ & $2.97_{-0.69}^{+0.97}$ & $0.34_{-0.11}^{+0.08}$  & -                       & -                       & -                      & -                       \\
            & 2007    & -                       & -                       & -                      & -                       & -                       & -                       & -                      & -                       \\
            & 2010    & $-1.23_{-0.12}^{+0.16}$ & $-3.64_{-0.14}^{+0.10}$ & $3.91_{-0.60}^{+0.82}$ & $0.50_{-0.04}^{+0.04}$  & $-1.00_{-0.10}^{+0.11}$ & $-3.47_{-0.07}^{+0.07}$ & $9.97_{-4.95}^{+6.60}$ & $2.06_{-0.03}^{+0.03}$  \\
            NGC 4051    & 2001    & $-1.17_{-0.14}^{+0.18}$ & $-3.61_{-0.24}^{+0.18}$ & $2.32_{-0.13}^{+0.14}$ & $-0.83_{-0.07}^{+0.06}$ & -                       & -                       & -                      & -                       \\
            & 2018    & $-1.69_{-0.09}^{+0.11}$ & $-2.96_{-0.12}^{+0.09}$ & $3.58_{-0.55}^{+0.73}$ & $-0.31_{-0.04}^{+0.04}$ & $-2.00_{-0.09}^{+0.13}$ & $-2.80_{-0.19}^{+0.12}$ & $4.81_{-2.09}^{+5.49}$ & $-0.39_{-0.03}^{+0.03}$ \\ \hline
            \end{tabular}
		\end{center}
	\end{table*}

\end{CJK*}	
\end{document}